\documentclass[aps,prl,groupedaddress,superscriptaddress,twocolumn,showpacs]{revtex4}

\usepackage{ae,aecompl}
\usepackage{amssymb}
\usepackage{amsmath}
\usepackage{helvet}

\usepackage[T1]{fontenc}
\usepackage[latin9]{inputenc}
\usepackage{graphicx}
\usepackage{epstopdf}
\usepackage{esint}
\usepackage{color}

\graphicspath{{Figures/}}

\newcommand{\bg}{ \begin{gather} }
\newcommand{\eg}{\end{gather}}
\newcommand{\be}{ \begin{equation} }
\newcommand{\ee}{\end{equation}}
\newcommand{\bea}{ \begin{eqnarray} }
\newcommand{\eea}{\end{eqnarray}}

\renewcommand{\Re}{\mathop{\rm Re}}

\begin{document}

\title{Strange metal state near quantum superconductor-metal transition in thin films }

\author{Konstantin S. Tikhonov}

\affiliation{L. D. Landau Institute for Theoretical Physics, Chernogolovka, 142432,
Moscow region, Russia}

\affiliation{Skolkovo Institute of Science and Technology, Moscow 143026, Russia}

\author{Mikhail V. Feigel'man}

\affiliation{L. D. Landau Institute for Theoretical Physics, Chernogolovka, 142432,
Moscow region, Russia}

\affiliation{Skolkovo Institute of Science and Technology, Moscow 143026, Russia}

\begin{abstract}
We develop a theory of quantum $T = 0$ phase transition (q--SMT) between metal and superconducting ground states in a two-dimensional metal with frozen-in spatial fluctuations $\delta\lambda(r)$ of the Cooper attraction constant. When strength of fluctuations $\delta\lambda(r)$ exceeds some critical magnitude, usual mean-field-like scenario of the q--SMT breaks down due to spontaneous formation of local
droplets of superconducting phase. The density of these droplets grows exponentially with the increase of  average attraction constant $\lambda$. Interaction between the droplet's order parameters is due to proximity effect via normal metal and scales with distance $\propto 1/r^\beta$ , with $2 < \beta \le 3$. We account for this interaction by means of a real-space strong-disorder renormalization group (RG). Near the q--SMT the RG flow is, formally, a dual equivalent of the Kosterlitz-Thouless RG. The corresponding line of fixed points describes a Griffiths phase of a metal with large fractal clusters of superconducting islands. Typical number of islands 
in a cluster grows as $N_\delta \sim 1/\delta$, where $0 < \delta \ll 1 $ is the distance to the critical point.
 Superconducting side is described by a runaway of RG trajectories into the strong-coupling region. Close to the transition point on the SC side, $0<-\delta \ll 1$, RG trajectories possess an extremum as function of the RG parameter $|\delta|^{1/2} \ln(1/T\tau)$. It results in a wide temperature range where physical properties are nearly $T$--independent. This observation may be relevant to the understanding of a \emph{strange metal} state frequently observed near q--SMT.
\end{abstract}

\date{\today}

\maketitle


\section{Introduction}
\label{sec:intro}
A number of potentially superconducting materials and alloys  lose their superconducting properties
upon increase of  disorder-induced electron scattering, suppressing superconducting transition temperature $T_c$ to zero.
The resulting state may be  either insulating, with a metallic state right at the quantum critical point,
or metal--like in the whole range of parameters.  The first situation is referred to as
Superconductor-Insulator Transition (SIT)\cite{goldman1998superconductor,gantmakher2010superconductor},
 while the second one as 
Superconductor-Metal Transition (SMT). 
 In this paper we will study SMT in two--dimensional (2D) or
quasi--2D disordered materials. 

It is usually assumed that a genuine metallic state cannot exist
in 2D  due to Anderson localization.  However,  the corresponding localization length is exponentially long
for not too strong disorder: $L_{\textrm{loc}}^{2D} \sim l e^{\pi g/2}$  where $g = h/e^2 R_\Box$ 
is the dimensionless conductance of the film at  high temperatures (when semiclassical Drude law is applicable)
 and $l$ is the elastic mean free path for electrons.
The corresponding energy (and temperature) scale where localization becomes relevant equals $T_{\textrm{loc}} \sim \tau^{-1} \exp(- \pi g)$, where $\tau$ stays for elastic scattering time.
  Below we consider situations when $g \gg 1$ is sufficiently large 
and thus exponentially low $T_{\textrm{loc}}$ can be treated as zero, since all temperatures in our problem
will be much higher. Under such an assumption, it is legitimate to consider  $T=0$  metal state and 
quantum phase transition of the SMT type in 2D (which may become SIT transition if the temperature is reduced below $T_{\textrm{loc}}$\cite{burmistrov2015superconductor}).

A natural mechanism of SMT transition upon increase of potential disorder is due to increase
of effective Coulomb repulsion between slowly diffusing electrons~\cite{Finkelstein1987,Finkelstein1994}.
The corresponding critial value $g_c$ of the Drude conductance $g$ equals 
$g_c = \frac1{2\pi}\ln^2\frac1{T_{c0}\tau}$ (with $T_{c0}$ for the 
 superconducting transition temperature of the same material in the clean limit), and can be rather large, $g_c \approx 10$. This mechanism is known to describe quite well the major features
of the SMT  in  a number of materials with high electron density and high disorder, like amorphous
 Mo$_x$Ge$_{1-x}$, Nb$_x$Si$_{1-x}$  and many other.  An extension of the Finkel'stein theory~\cite{Finkelstein1987,Finkelstein1994} was developed to treat inhomogeneous systems composed of small 
superconducting islands in contact with dirty metal~\cite{FL1998,SLF2001,SpivakZuzinHruska}.
Such an approach allows to locate the SMT position depending on the system parameters (conductance $g$, fraction
of superconducting regions $x \ll 1$, etc.)

However, the nature of the ensuing metal phase
realized at $g < g_c$ at very low temperatures is not understood yet. 
Strong enhancement of conductance (compared to  its magnitude in the normal state $g$) is frequently 
observed~\cite{KKS} in the vicinity of a quantum transition to superconducting
state. Surprisingly, conductance is weakly $T$--dependent in this phase dubbed therefore `strange metal'.
In some cases~\cite{Tamir19},  strange metal state has been shown to have extrinsic origin 
(insufficient filtering of high-frequency noise in the measuring system). However, it is not clear if all numerous observations of a `strange metal'  state are of the same origin.
In the present paper we discuss  another possible origin of the strange metal behavior: a $T=0$ 
Griffiths phase dominated by large statistical fluctuations due to frozen--in fluctuations of the Cooper interaction amplitude. 

To derive these new results,
we employ a model of  a diffusive metal with relatively large conductance $g$ and spatially fluctuating
Cooper interaction $\lambda(\mathbf{r}) =  \bar{\lambda} + \delta\lambda(\mathbf{r})$ ($\lambda>0$ corresponds to attraction) . We find that at sufficiently strong disorder $w$ (to be defined precisely below) an unusual localization transition responsible for a quantum SMT occurs upon increase
of the mean value of the Cooper interaction $\bar{\lambda}$. 

The rest of the paper is organized as follows.  In the Sec. \ref{sec:AL} we formulate and study a special kind of
Anderson localization problem that describes  eigenvalue spectrum and eigenfunctions of the propagator of
superconducting fluctuations $L(\mathbf{r}, \mathbf{r'})$ within the random--$\lambda$ model.   We present numerical results  for the spectrum density $\rho(E)$ 
and statistical properties of the eigenfunctions $\psi_E(\mathbf{r})$ related to the $L(\mathbf{r}, \mathbf{r'})$. 
 It will be shown that eigenfunctions near the lower edge of the spectrum $E_0$ (defined as $\rho(E) = 0$ at $E < E_0$) 
 are extended at relatively small values of dispersion $w < w_c$, but  localized at 
 $w > w_c$,  with localization length being rather short  close to the transition point
$w=w_c$. As we mention in the conclusion, the strong-disorder regime $w > w_c$
is easily realized upon approach to the quantum SMT.  Thus the nature of q--SMT is determined
by  \textit{emergent superconducting granularity}. Note the difference between this phenomenon and previously studied 
superconducting granularity that is due to strong random potential scattering of electrons~\cite{Ghosal01,Feigelman10a,Feigelman10b,Bouadim11,stosiek2019self} and occurs near SIT. In the model we discuss in the present paper superconducting islands
occur within a `sea' of a normal metal.

In  Sec. \ref{sec:islands} we extend our analysis to non--linear and non--local terms of the dynamic 
Ginzburg--Landau action. We study phase dynamics of individual superconducting islands
 and derive an effective  interaction between phases of \textit{different} localized islands. 
We show that this interaction is of the same functional form that is known for proximity-induced 
Josephson coupling between artificially prepared superconducting islands on top of diffusive 2D metal,
analyzed previously  in Refs.~\cite{FL1998,SLF2001,SpivakZuzinHruska,STF2008}.  Upon increase of
mean Cooper attraction $\bar{\lambda}$, this interaction becomes strong enough to produce 
correlations between phases of different islands. At this stage,  a macroscopic description
of superconducting correlations on a length scales containing many original islands becomes necessary.

To treat these correlations quantitatively,  in Sec. \ref{sec:RG} we use a version of
Strong-Disorder Renormalization Group (SDRG),  originally due to 
D. S. Fisher~\cite{Fisher92,Fisher95}, and extensively reviewed in~\cite{Igloi-Monthus05,Igloi-Monthus14,
Refael13}.  We find our problem to be formally similar to the one studied in Ref.~\cite{Igloi14}
and identify disordered Griffiths phase with a line of fixed points of the SDRG transformations.
  Long-range interaction between localized islands of superconductivity leads to formation of 
strongly coupled fractal clusters of islands  with a slow collective dynamics. Superconducting phase
is then identified with a runaway of the SDRG solution into the strong-coupling regime and generation
of a long-but-finite spacial scale where macroscopic superconducting coherence sets in. In Sec.~\ref{sec:SM} we discuss low-temperature physics of the strange metal and superconducting phases. Finally, Sec.~\ref{sec:Conclu} is devoted to the discussion of results and conclusions. Supplemental material Sections (S1 and S2)
contain a number of  technical details  of our theory.


\section{Anderson localization of superconducting modes}
\label{sec:AL}
\subsection{Model}
We consider a model of normal metal with moderately large dimensionless conductance $g \gg 1$,
with phonon-mediated Cooper attraction characterized by BCS coupling strength $\lambda_*$.
Coulomb interaction between electrons is considered to be in the `universal limit', i.e. 
screened static Coulomb potential is equal to $\nu^{-1}$, where $\nu$ is the electron density of
states (per single spin projection). The effect of Coulomb interaction and disorder upon superconducting
instability threshold~\cite{Finkelstein1987,Finkelstein1994} can be represented~\cite{FLS2000,SLF2001}
via effective repulsion constant $\lambda_g = 1/\sqrt{2\pi g} \ll 1$. It is important to notice that
effective repulsion $\lambda_g$ cannot be simply subtracted from the attraction constant $\lambda_*$.
Indeed,  superconducting instability due to the presence of $\lambda_*$ can be described by summation
of ladder diagrams within Cooper channel only (electron processes with a small total momentum of electron pair).
On the other hand, Coulomb interaction  enters via the density-density channel, and it is
necessary to take into account non-ladder diagrams of `parquet' type. In the mean-field scenario, quantum phase transition from metal
to superconducting state occurs upon increase of the average attraction $\langle \lambda_* \rangle $ 
against the background of  repulsion those strength is determined by $\lambda_g$.

The stability of the normal state with respect to the effect of superconducting inclusions is determined by the properties
of superconducting  propagator $L(\omega, q)$. 
The quadratic part of the Ginzburg--Landau (GL) functional reads
\be
S_2=\int \frac{d\omega}{2\pi} \int d\mathbf{r}_1 d\mathbf{r}_2 
\Delta_{\omega} (\mathbf{r}_1)\Pi(\omega,\mathbf{r}_1-\mathbf{r}_2)\Delta_{\omega}(\mathbf{r}_2).
\label{S2}
\ee

We start with expression for $\Pi(\omega,\mathbf{q})$  at $\omega=0$ and in  homogeneous system. 
Technically it is convenient to account for the effect of Coulomb repulsion on the Cooper channel 
in the way it was done in Refs. \cite{Oreg1999,Skvortsov2005}. Namely, we introduce the
Cooperon screening factor $w_q(\epsilon)$ which modifies usual expression for the Cooperon amplitude
$C_q(\epsilon) = \nu/(Dq^2 + 2|\epsilon|)$ multiplicatively: $C_q(\epsilon) \to C_q(\epsilon) w_q(\epsilon)$.
In the limit of $T\to 0$, the screening factor $w_q(\epsilon)$ obeys then the following equation:
\be
w_q(\epsilon) = 1 - \int \frac{d\epsilon_1}{2\pi} 
\frac{2\theta(\epsilon\epsilon_1)}{g}\ln\frac1{|\epsilon+\epsilon_1|\tau}\frac{ w_q(\epsilon_1)}{Dq^2 + |\epsilon_1|},
\label{wq0}
\ee
where $\theta(x)$ is the Heaviside step function.
Equation (\ref{wq0}) for the function $w_q(\epsilon) \equiv w_q(\zeta)$, with $\zeta=\ln(1/\epsilon\tau)$,
can be rewritten (within the  logarithmic accuracy) in the following form:
\be
w_q(\epsilon) = 1 - \lambda_g^2 \int_0^{\zeta_q} d\zeta_1 \min(\zeta,\zeta_1) w_q(\zeta_1),
\label{wq}
\ee
where $\zeta_q= - 2\ln(ql)$.  Solution of Eq.~(\ref{wq}) with initial condition
$w_q(0)=1$  is
\begin{eqnarray}
\label{wq2}
w_q(\zeta \leq \zeta_q) & = & \cosh(\lambda_g\zeta) - \tanh(\lambda_g\zeta_q)\sinh(\lambda_g\zeta),  \\ \nonumber
w_q(\zeta \geq \zeta_q) & = & \frac1{\cosh(\lambda_g\zeta_q)}.
\end{eqnarray}
The Cooperon screening factor modifies $\Pi(0,\mathbf{q})$ as follows:
\be
\Pi(0,\mathbf{q})=\frac{\nu}{\lambda _{\ast }} - \int d\epsilon \frac{\nu }{Dq^{2}+2\left\vert\epsilon \right\vert }w_q(\epsilon).
\label{a1}
\ee
Substitution of Eq.~(\ref{wq2}) into Eq.~(\ref{a1}) and integration leads to (we also account for finite frequency $\omega$)
\be
\Pi(\omega,\mathbf{q})=\frac{\nu}{\lambda _{\ast }}-\frac{\nu}{\lambda _{g}}+
\frac{\nu}{\lambda _{g}}\Pi_0(\omega,\mathbf{q}),
\label{a2}
\ee
with 
\be
\label{Cdef}
\Pi_0(\omega,\mathbf{q})=\frac{2}{1+ \left[(ql)^2 + 2\omega\tau \right]^{-2\lambda _{g}}},
\ee 
which at $ql\ll 1,\;\omega\tau\ll 1$ becomes $\Pi_0(\omega,\mathbf{q}) \approx 2\left[(ql)^2 + 2\omega\tau \right]^{2\lambda_g}$.
Now  we account for spatial fluctuations of the bare Cooper attraction constant $\lambda_*$
and replace the function $\Pi(\omega,\mathbf{q})$  by the operator
\be
\hat{\Pi}(\omega,\mathbf{q};\mathbf{r})=\frac{\nu}{\lambda_g}\left[\frac{\lambda_g}{\lambda_*(\mathbf{r})}-1+\Pi_0(\omega,\mathbf{q})\right]
\label{a3}
\ee
  It is more convenient to parametrize the disorder as
\be
\frac{\lambda_g}{\lambda_*(\mathbf{r})}-1 = \delta_0 + u(\mathbf{r}),
\label{a4}
\ee
assuming that random field $u(\mathbf{r})$  has zero mean and 
\be
\label{defd0}
\delta_0 =\left<\frac{\lambda_g}{\lambda_*(r)}\right>-1
\ee is the bare distance to
superconductor-metal transition at $T=0$. We will assume $u(\mathbf{r})$ to be Gaussian with correlation function 
$ \overline{u(\mathbf{r}) u(\mathbf{r}')} = \Lambda f(|\mathbf{r}|/b)$ (although by definition $u(\mathbf{r})\ge-1-\delta_0$, this constraint is not problematic, as we discuss below). Here $\Lambda$ is the dimensionless
fluctuation strength and $b$ is the correlation length, while function $f(x)$ is assumed to be 
fast decaying at $ x \geq 1$ and normalized according to $\int f(\mathbf{x}) d^2x = 1$.
It turns out the properties of $L(r)$ depend crucially on the value of 
dimensionless parameter $w = \Lambda b^2/l^2$, which is determined by \textit{both} 
 the fluctuations of $\lambda_*(\mathbf{r})$ and the electronic mean free path $l$.

The superconducting propagator $L(\omega, \mathbf{q})$ is given by the Fourier transform of the average solution 
$\mathcal{L}(\omega; \mathbf{r},\mathbf{r}^{\prime})$ of the equation
\begin{eqnarray}
\nonumber
\left(\delta_0 +u(\mathbf{r})\right)\mathcal{L}(\omega; \mathbf{r},\mathbf{r}^{\prime}) +
\int d^{2} \mathbf{r}_{1} 
\Pi_0(\omega,\mathbf{r}-\mathbf{r}_{1})
\mathcal{L}(\omega; \mathbf{r}_{1}, \mathbf{r}^{\prime})  \\ 
= \frac{\lambda_g}{\nu}\delta(\mathbf{r}-\mathbf{r}^{\prime}). \,\quad
\label{D1}
\end{eqnarray}

\begin{figure*}[tbp]
\minipage{0.33\textwidth}\includegraphics[width=\textwidth]{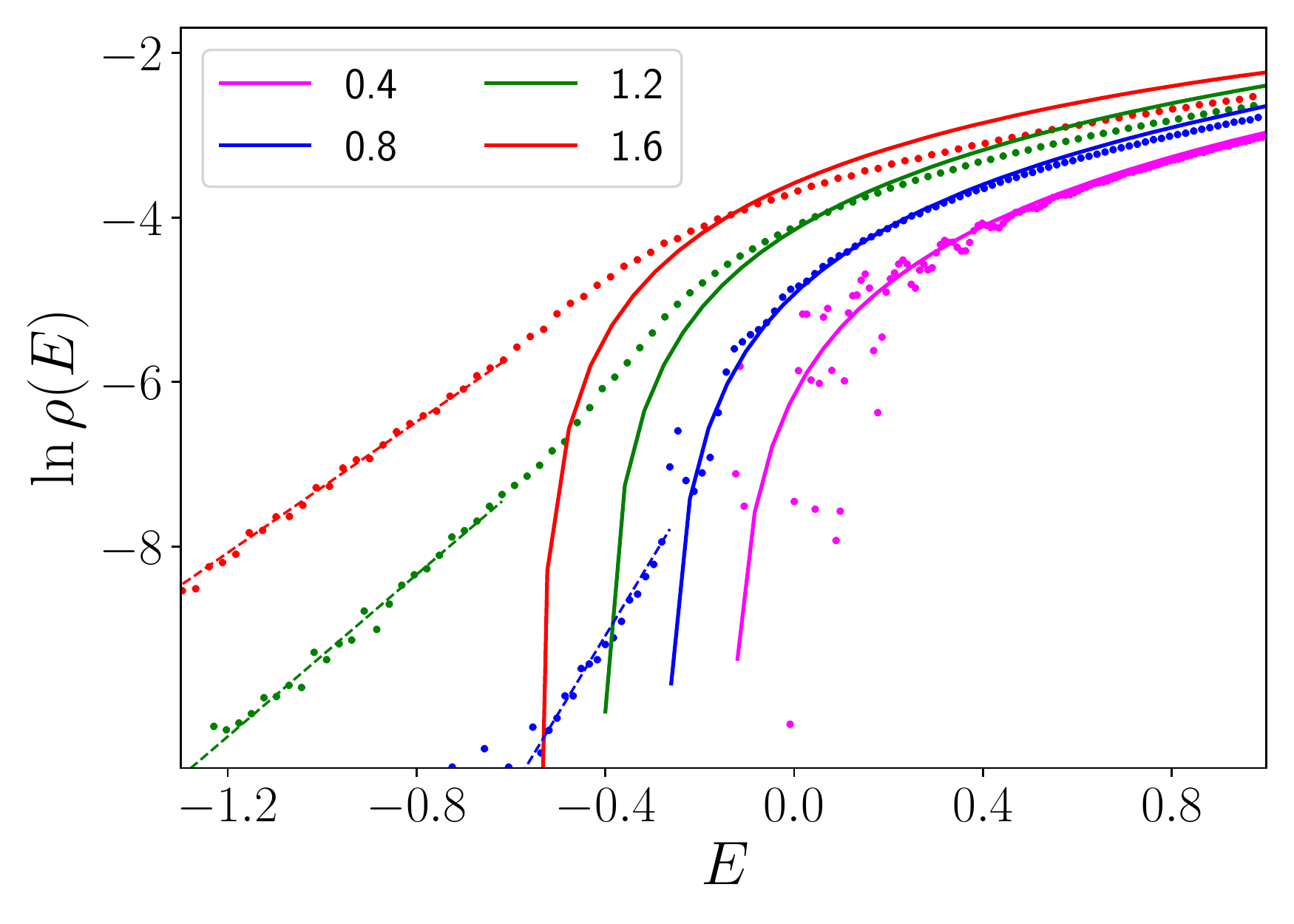}\endminipage
\minipage{0.33\textwidth}\includegraphics[width=\textwidth]{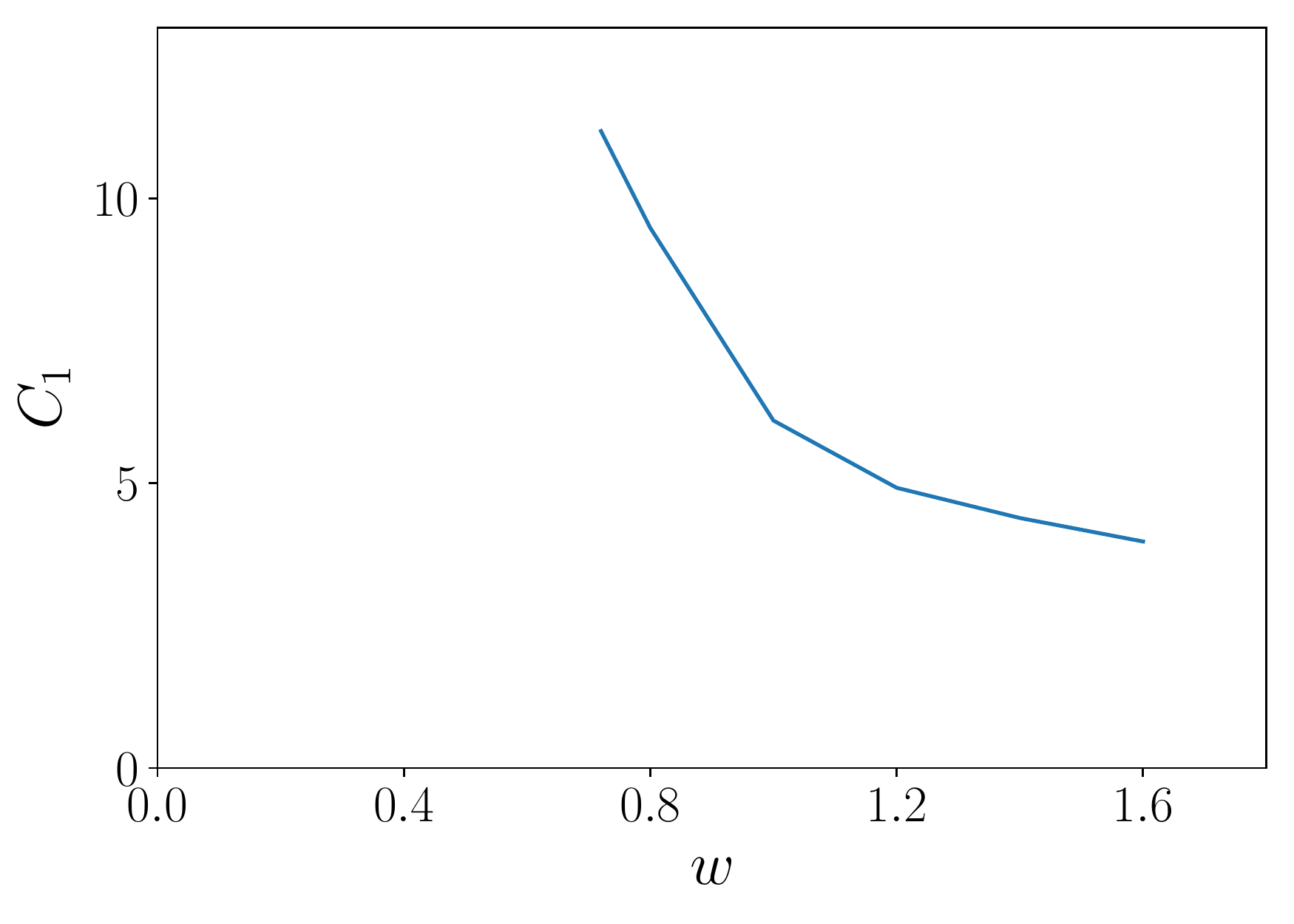}\endminipage
\minipage{0.33\textwidth}\includegraphics[width=\textwidth]{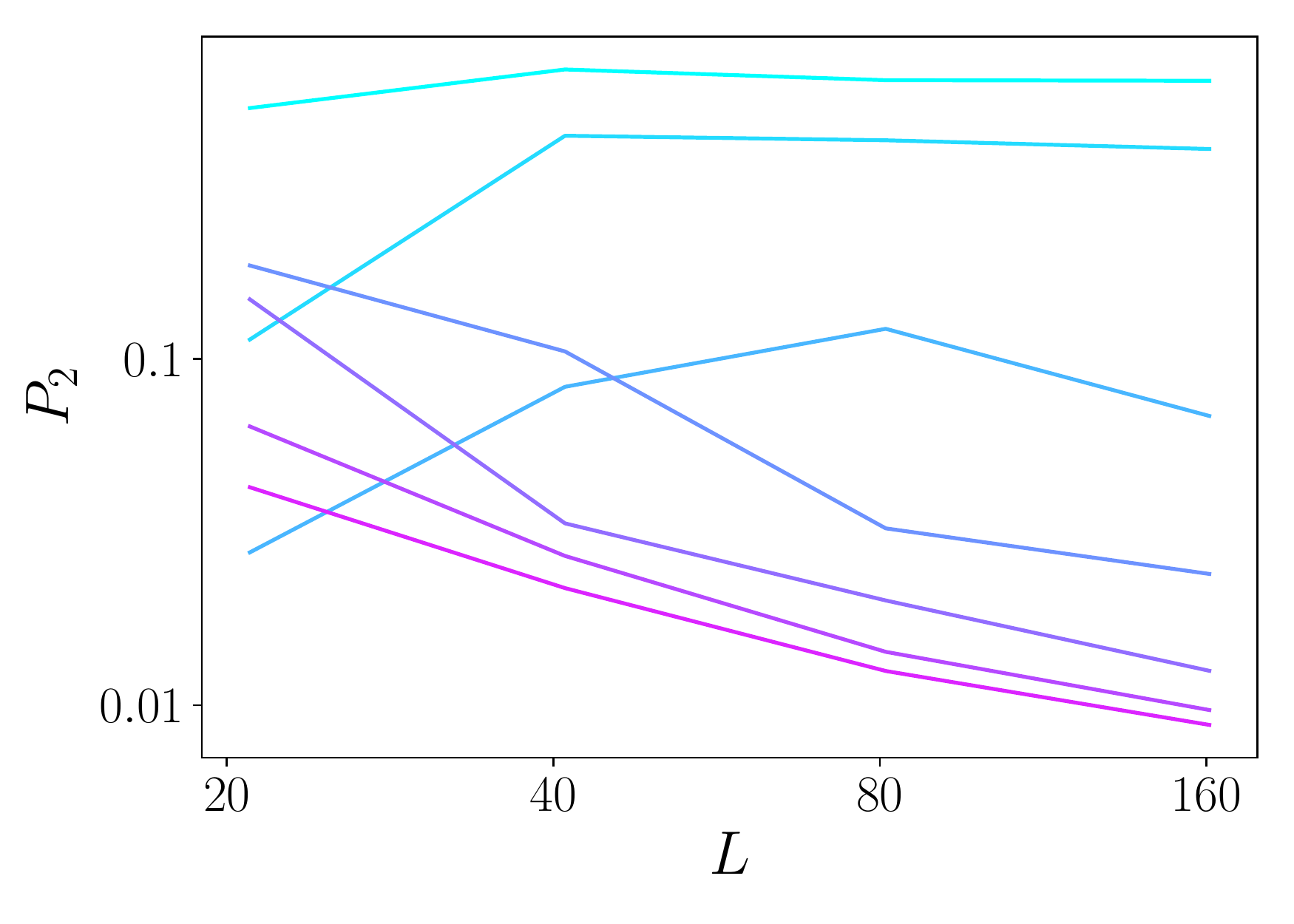}\endminipage
\caption{Characterization of the eigenstates of Eq.~(\ref{E1}) at $\lambda_g=0.2$ and several values of disorder $w$. a: DoS $\rho(E)$ found from ED (dots), disorder $w$ is indicated on the legend. Solid lines: $\rho(E)$ as found from SCBA, Eqs (\ref{dosscba}) and (\ref{eqscba}). In the tail: dashed lines are fits by Eq.~(\ref{logdos}). b: $w$-dependence of the coefficient $C_1$ in Eq.~(\ref{logdos}). c: Inverse participation ratio $P_2$ at $w=1.2$ for several energies $E=-0.8, -0.6, -0.4, -0.2, 0.0, 0.2, 0.4$ from cyan to magenta. Observe that states become less localized upon increasing the energy $E$ from the tail to the bulk of the spectrum.}
\label{fig:pg}
\end{figure*}

As a result,

\begin{equation}
 L_{\omega}(\mathbf{r}-\mathbf{r}') = \overline{\mathcal{L}(\omega, \mathbf{r}, \mathbf{r}^{\prime})}  = \frac{\lambda_g}{\nu} \overline{\sum_{n}\frac{\psi_{n}(\mathbf{r})\psi_{n}^{*}(\mathbf{r}^{\prime})}{E_{n}+\delta_0 - i0}},
\label{LR1}
\end{equation}
where $\psi_n(\mathbf{r})$ and $E_n$ are determined by the following equation:
\begin{widetext}
\begin{equation}
\label{E1}
\int d^{2} \mathbf{r}_{1} [\Pi_0(\omega, \mathbf{r}-\mathbf{r}_{1}) + u(\mathbf{r})\delta(\mathbf{r}-\mathbf{r}_{1})]
\psi_n(\mathbf{r}_{1};\omega) = E_n\psi_n(\mathbf{r};\omega).
\end{equation}
\end{widetext}
To avoid confusion, we emphasize that variable $E$ in the above equation (and below)
has nothing to do with single-electron energies.

The critical point of the mean-field transition is determined by the divergence of
$L(\omega\to0, q \to0)$. In the absence of Cooper constant fluctuations, $w=0$, it occurs at $\delta_0=0$.
Relatively weak fluctuations $u(\mathbf{r})$ shift it to some nonzero $\delta_c$, as long as
eigenfunctions $\psi_n(\mathbf{r})$ remain delocalized near the band edge, i.e. for smallest values of
eigenvalues $E_n$ of the operator (\ref{E1}) at $\omega=0$.
In this case $\delta_c = -\min(E_n)$.  Below we find, at sufficient increase of $w$, 
localization of eigenfunctions with eigenvalues close to the spectral edge, invalidating 
this simplest mean-field scenario.

\subsection{Numerical analysis}
We start from evaluation of  the Density of States (DoS)  of the operator defined by Eq.~(\ref{E1}),
 in the self-consistent Born approximation (SCBA). DoS   $\rho(E)$  is then determined by the following equation:
\begin{equation}
\label{dosscba}
\rho(E)=\frac{1}{4\pi^2}\textrm{Im}\Sigma(E+\sigma(E)),
\end{equation}
where $\sigma(E)$ can be found from nonlinear self-consistency equations
\begin{eqnarray}
\label{eqscba}
\sigma(E)=w^2\Sigma(E+\sigma(E)),\;\;\; \\ \nonumber
\Sigma(E)=\frac{1}{2\pi}\int_0^{1}\frac{(ql) d(ql)}{-E + \Pi_0(0,q)}.
\end{eqnarray}
Notice that disorder enters Eq.~(\ref{eqscba}) via the parameter $w = \Lambda b^2/l^2$, where 
factor $1/l^2$ appears due to the high-momentum cutoff  $q_{\textrm{max}} = 1/l$ in the integral. 
Solving  Eq.~(\ref{eqscba}) numerically and evaluating the DoS in Eq.~(\ref{dosscba}), we find the
 results, presented on the Fig. \ref{fig:pg}a by solid lines for several values of disorder.
 According to these results, the sharp edge of the eigenvalue spectrum
survives within SCBA  approximation, and the position of this edge is $-\delta_c(w)$. For $\lambda_g=0.2$, the function $\delta_c(w)$ was determined numerically to be $\delta_c(w)=0.34 w$ 
in the range of $w \lesssim 1.6$; in general, the slope $d\delta_c/dw$
 depends on the value of $\lambda_g$.
  
Thus the SCBA result for the average superconducting propagator 
reads, in the infrared limit $ql, \omega\tau \ll 1$:
\be
L_\omega(q) = \frac{\lambda_g}{\nu}\frac{1}{\delta_{\textrm{SCBA}} + 2 [(ql)^2 + 2|\omega|\tau]^{2\lambda_g} },
\label{L2}
\ee
where $\delta_{\textrm{SCBA}} = \delta_0 - \delta_c$ is the distance to the critical point  determined within SCBA.
Note that the primary effect of fluctuations in $1/\lambda_*(\mathbf{r})$ is to strengthen a tendency 
to superconducting instability, which occurs now at some positive $\delta_0 = \delta_c(w)$.

To check the above results, we evaluate the same DoS via exact diagonalization (ED) of the discretized operator in Eq.~(\ref{E1}), with the lattice constant equal to the mean free path $l$.  For disorder, we choose correlated Gaussian distribution with correlation function 
\be
\left<u(\mathbf{r})u(0)\right>=\frac{\Lambda}{2\pi}K_0(r/b)
\label{u-corr}
\ee
at long distances $r\gg l$.

 We discuss here the case of short-range correlations, $b=0.25$. 
The resulting $\nu(\epsilon)$ in a broad energy range is shown on the Fig. \ref{fig:pg}a.  
We are mainly interested in the properties of the spectrum 
at energies at and below the spectrum edge of an ideal ($w=0$) system. The results for  $\rho(E)$ found from ED are shown in the main panel of Fig. \ref{fig:pg}a with dots for several values of disorder for square lattice of linear size 
$L=161$ with periodic boundary conditions.  Apparently, the SCBA describes the exact DoS well for large enough energy $E \gtrsim E_*(w)$ for all disorder strengths $w$.  The same is true for smallest disorder 
$w=0.4$ in the whole range of energies: the effect of disorder reduces to the shift of the spectrum edge
by $-\delta_c$, see Fig.~\ref{fig:pg}a. 

However, at slightly stronger disorder, $w=0.8$, an
enhancement (with respect to SCBA result) of the DoS at negative $E$ is already seen; the same feature
 becomes more evident at larger disorder, $w=1.2-1.6$. Functional form of this tail fits well by simple
exponential dependence
\be
\label{logdos}
\ln\rho(E) = C_1(w)E - C_0(w).
\ee
Parameter $C_1$ as function of disorder strength $w$ is shown in Fig.~\ref{fig:pg}b. Exponential form (\ref{logdos}) of the DoS will play a 
crucial role in our analysis below. 
 Before proceeding, let us stress that the fact that the left tail of the DoS $\rho(E)$ extends to arbitrary large negative $E$ is related to the assumption of Gaussianity of $u(\mathbf{r})$. Due to the constraint, mentioned after Eq. (\ref{defd0}), the Gaussian approximation fails for $E\approx E_{\min}=-1-\delta_0$ and a sharp band edge should be present at $E= E_{\min}$ in any realistic model for $u(\mathbf{r})$. This is not a problem as there are still plenty of states for which the Gaussian approximation is valid.

In order to characterize the wavefunctions in various parts of the spectrum, we calculate the inverse participation ratio $P_2=\left<\sum_r \psi^4(r)\right>$ (averaging over disorder realizations is implied) as function of the system area $S=L^2$ and energy $E$ at several $w$. Generally, one expects this scaling to be of the power-law form $P_2(S\to\infty)=S^{-\mu}$ with $\mu$ distinguishing between metallic ($\mu=1$), insulating ($\mu=0$) or fractal (anything in between) behavior of the wavefunctions.  Fig. \ref{fig:pg}c illustrates that the states become much less localized with increase of $E$.

Our problem belongs to a class of problems with determenistic power-law hopping and on-site disorder \cite{levitov1989absence,Burin89,levitov1990delocalization}, see a recent review \cite{syzranov2018high}. The data discussed above indicate that the eigenfunction, corresponding to the lowest eigenvalue of the operator defined in Eq.~(\ref{E1}), undergoes a localization transition that happens  with increase of the disorder. Similar transition was studied earlier in~\cite{Malyshev1,Malyshev2,Malyshev3}  
for other power-law-tunneling models with long-range tunnelling amplitude decaying as $r^{-\beta}$ and
 on-site disorder in 1D and 2D cases; see also recent 
papers~\cite{KravtsovPower1,KravtsovPower2,deng2020anisotropymediated}. Although it is expected that all states (that is, bulk states together with the edge one) are localized in thermodynamic limit at $w>w_c$ and $d=2$ in such a 
system~\cite{deng2020anisotropymediated}, we will make use of the fact that localization length of the excited states with sufficiently large $E$ becomes long, as demonstrated in Fig.~\ref{fig:pg}c.

 Perturbative result about the absence of localization of
the edge modes  at \textit{weak disorder} follows then from simple power-counting arguments provided
in Ref.~\cite{Malyshev2}.  In our model  real-space hopping $ \sim r^{-\beta}$ originates from
non-analytic behavior of the kernel given in Eq. (\ref{Cdef}),
thus $\beta = 2 + 4\lambda_g$ in our case. More precisely, in Eq. (\ref{E1}) we have (at small $\lambda_g$):
\begin{equation}
\Pi_0(0,\mathbf{r})\approx -\frac{4\lambda_g}{\pi l^2} |\mathbf{r}/l|^{-\beta}
\end{equation} at $|\mathbf{r}|\gg l$. For 2D space, the arguments of Ref.~\cite{Malyshev2} are
 valid  under the condition $ 2 < \beta < 3$, thus the above inequality translates to  $0 < \lambda_g < 1/4$.

With increase of  disorder $w$ the edge localization transition occurs, 
leading to appearance of localized eigenstates in the Lifshitz tail. 
In the present paper, we do not aim to study this specific  transition in details.
Our analysis in what follows will rely on the appearance of the well-defined exponential tail of the spectrum, 
with  localized eigenstates,  at super-critical disorder $w > w_c(\lambda_g)$.
For $\lambda_g \geq 1/4$, localized states in the tail appear at any disorder and $w_c=0$.

The papers~\cite{suslov1994density,syzranov2015unconventional} predict for this type of problems (where usual smooth solution for Lifshits tail does not exist)
 existence of a Gaussian tail in the DoS, $\ln\rho(E)\propto-E^2$, for \textit{arbitrary weak} disorder.
This refers to the usual Schrodinger equation with random potential in high dimensions $d > 4$~\cite{suslov1994density},
 and to systems with a power--law quasiparticle dispersion~\cite{syzranov2015unconventional} of the type we consider at $\lambda_g < d/8$. Our numerical data (not shown) provide signatures of existence of this kind of states at weak disorder, although with extremely low values of DoS, much smaller than in a
simple exponential tail shown in Fig.~\ref{fig:pg}a for a super-critical disorder.

To conclude this Section, we emphasize again the dependence of the key parameter $w = \Lambda b^2/l^2$ 
on both spatial fluctuations  of Cooper constant (parameters $\Lambda$ and $b$) and on electronic  mean free path $l$. 
In brief,  $\delta\lambda(\mathbf{r})$ fluctuations are more efficient when elastic scattering is strong.


\section{Localized superconducting islands}
\label{sec:islands}

We have found in Sec. \ref{sec:AL}  that eigenvalue spectrum  for static superconducting fluctuations 
$\Delta(\mathbf{r})$, as determined by Eq.~(\ref{D1})  at $\omega=0$, 
is unbounded from below once disorder parameter is super-critical, $w > w_c$. 
 According to Eq.~(\ref{LR1}), it leads to instability of all fluctuation modes 
with eigenvalues  $E_n < -\delta_0$.  Number of these (linearly) unstable
modes grows fast upon increase of average Cooper attraction $\bar{\lambda}$.  
The amplitudes of these modes become finite upon account of nonlinear terms in the action,
so one finds emergent  superconducting `islands' immersed in a normal metal.

Thus our strategy is, first, to study the properties of  localized islands of superconductivity,
and in particular  dynamics of the order parameter phases $\varphi_i(t)$ associated with those islands.
Second, we will account for the interaction between phases of different islands.  Such
interaction comes about due to nonlinear coupling between localized and delocalized eigenmodes
of the linear problem, which are defined by Eq.~(\ref{E1}). Then the coupling between phases $\varphi_n$ and $\varphi_m$
of different localized modes  is mediated by the propagator of delocalized modes, Eq.~(\ref{L2}).

\subsection{Effective single-island action and its parameters}

Individual localized modes are described by order parameters $\Delta_i(\mathbf{r},t) = a_i(t) \psi_i(\mathbf{r})$ where 
$\psi_i(\mathbf{r})$ are normalized eigenfunctions of the linear problem (\ref{D1}), 
and $a_i(t)$  are time-dependent complex amplitudes. The imaginary-time action $S\left[a(t)\right]$ 
in terms of the amplitude $a_i(t)$ is:
\begin{equation}
S = \nu\left[ \int dt  \left(\frac{\alpha_i}{\lambda_g} |a_i|^2 + \frac{B_i}{2} |a_i|^4 \right) +
\int \frac{d\omega}{2\pi} \Gamma_i |\omega||a_i^{(\omega)}|^2 \right],
\label{action1}
\end{equation}
where $a_i^{(\omega)}$ is the Fourier-transformed $a_i(t)$.
Here $\alpha_i = E_i + \delta_0 $, so all modes with negative $\alpha_i$ are linearly unstable and the account of
quartic term is mandatory;  we  discuss the vertex $B_i$  a bit later. 
Last term in $S\left[a(t)\right]$ accounts for damping of superconducting fluctuations in spirit of time-dependent Ginzburg-Landau  (TDGL)  theory. 
It is non-local in the imaginary-time representation, so we prefer to present it in the frequency domain.
 Note that under the assumed condition
that superconducting islands cover small portion of system area, such a dissipative term is  natural,
as dissipation is provided by gapless electrons in surrounding metal.    Within usual TDGL theory operating at $T >0$,
the coefficient $\Gamma \sim 1/T$ and seems to diverge in the $T=0$ limit.  Indeed it is the case in the
standard scaling theory, Ref.~\cite{BK}.   The crucial point of our present analysis is that we deal here with
\textit{localized} superconducting fluctuations, and the presence of finite localization length $L_i$ corresponding
to an eigenmode $\psi_i(\mathbf{r})$, leads also to a finite value of the kinetic constant $\Gamma_i \sim L_i^2/D$, i.e.
it is given by electron diffusion time through the size of the corresponding superconducting eigenmode. 
The same estimate was obtained in Ref.~\cite{SpivakZuzinHruska} by a different method and for somewhat 
different formulation of the problem.

\begin{figure}[tbp]
\includegraphics[width=0.5\textwidth]{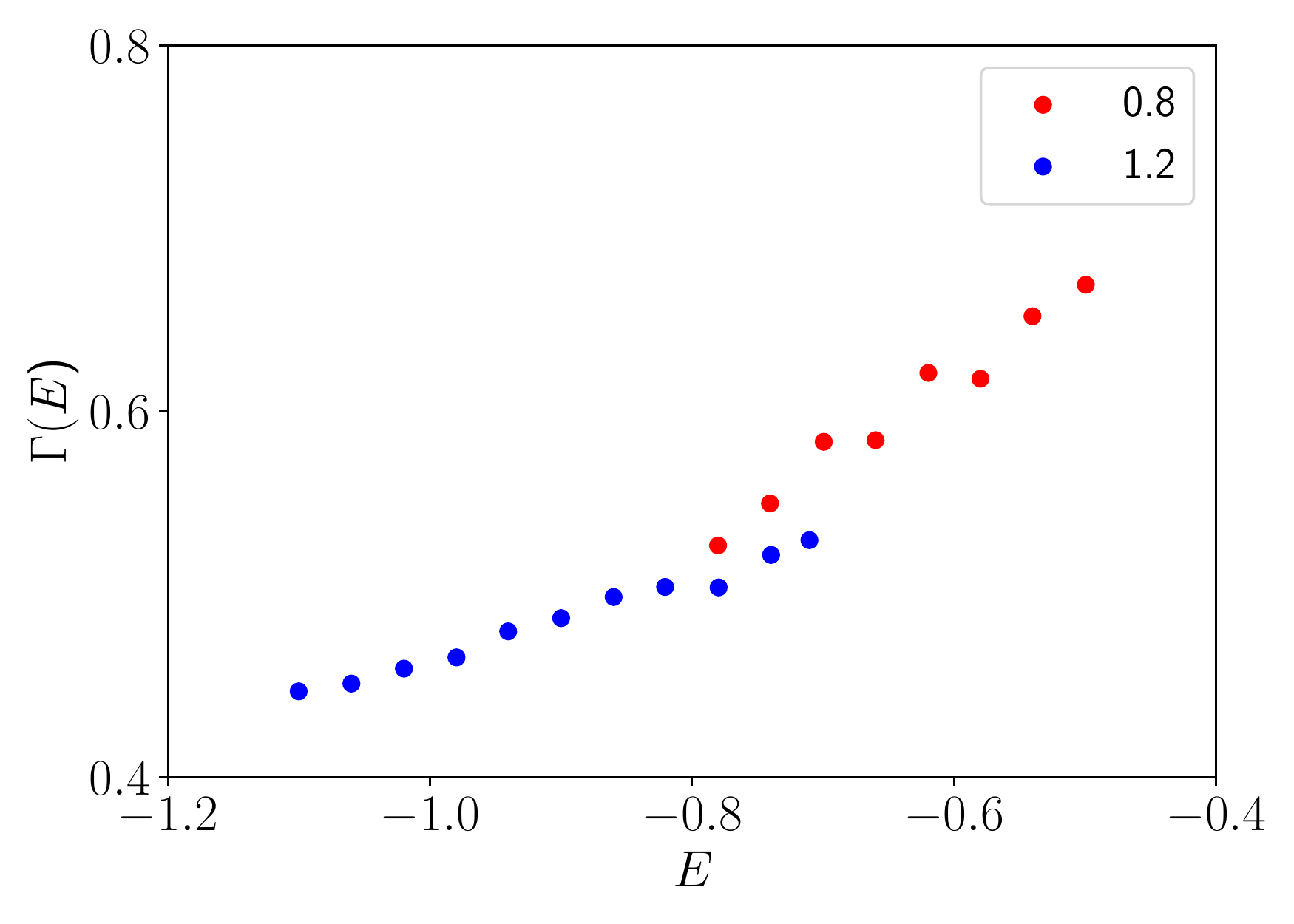}
\caption{The coefficient $\Gamma(E)$ in Eq.~(\ref{Eiomega}) for two disorder values $w=0.8$ and $w=1.2$, as shown in legend.}
\label{fig:ce}
\end{figure}

Quantitatively, the value of $\Gamma_i$ can be found by the analysis of the $\omega$ - dependence of an eigenvalue
$E_i(\omega)$ of the general linear operator defined by Eqs.~(\ref{Cdef},\ref{D1}) in the range of  small
$\omega \ll D/L_i^2$.  Indeed, at small $\omega$ the major part of this $\omega$-- dependence can be obtained
by a first-order perturbation theory over $\omega$, without modification of the eigenstate $\psi_i(\mathbf{r})$.
It leads then to linear in $|\omega|$  correction to the  eigenvalues 
\begin{equation}
\label{Eiomega}
E_i(\omega) = E_i(0) + \Gamma_i |\omega|
\end{equation}
 which leads to the last term of the action (\ref{action1}). The coefficient $\Gamma_i$ depends on energy $E_i$
and disorder $w$. Illustrative examples of such a dependence are shown in Fig.~\ref{fig:ce} for two different
strengths of disorder.

Now we turn to the estimates for the quartic vertex $B_i$.   Again, finite localization length $L_i$ leads to
a finite value of $B_i$  even in the $T=0$ limit, contrary to usual results~\cite{BK}.  Simple dimension estimates
indicate that $B_i \sim L_i^2/D^2$. 
For accurate derivation of $B_i$, we start from general expression for the quartic term in the 
action as functional of slowly varying order parameter field $\Delta(\mathbf{r},t)$,
\be
S_4[\Delta(\mathbf{r},t)] = \int dt F_4( \Delta(\mathbf{r},t)),
\label{S4}
\ee
where quartic part of the free energy $F_4(\Delta)$ acquires the form (for time-independent $\Delta(\mathbf{r})$):
\be
F_4(\Delta)= \frac{\nu}{2}\int\Pi_{i=1}^4 d^2r_i \Delta(\mathbf{r}_1)\Delta^*(\mathbf{r}_2)\Delta(\mathbf{r}_3)
\Delta^*(\mathbf{r}_4)
B_4(\left\{\mathbf{r}_i\right\}) 
\label{F4}
\ee
with (compare with Eq.~(15) in Ref.~\cite{Galitski03}):
\begin{widetext}
\bea
\label{B4}
B_4(\left\{\mathbf{r}_i\right\}) 
 =\pi T\sum_\epsilon\Pi_{k=1}^4\frac{w_k(\epsilon)}{|\epsilon|+\frac12 D (-i\partial_k)^2}
\delta(\mathbf{r}_1-\mathbf{r}_2)\delta(\mathbf{r}_1-\mathbf{r}_3)\delta(\mathbf{r}_1-\mathbf{r}_4)
  H_{\epsilon}(-i\partial_1,-i\partial_2,-i\partial_3,-i\partial_4)= \nonumber\\
  =\pi T\sum_\epsilon\Pi_{k=1}^4 \int \frac{(d^2\mathbf{p}_k) w_k(\epsilon) }{|\epsilon| + Dp_k^2/2} 
\delta(\mathbf{p}_1 + \mathbf{p}_3 - \mathbf{p}_2 - \mathbf{p}_4)
e^{i \mathbf{p}_1\mathbf{r}_1 + i \mathbf{p}_3\mathbf{r}_3  
- i \mathbf{p}_2\mathbf{r}_2 - i \mathbf{p}_4\mathbf{r}_4  }
  H_{\epsilon}(\mathbf{p}_1,\mathbf{p}_2,\mathbf{p}_3,\mathbf{p}_4),
\eea
\end{widetext}
where `screening factors' $w_k(\epsilon) \equiv w_{\mathbf{p}_k}(\epsilon)$ are given by Eqs.~(\ref{wq2}) and Hikami box reads
\be
H_{\epsilon}(\left\{\mathbf{p}_i\right\})=|\epsilon| + \frac{D}{8}\left[(\mathbf{p}_1-\mathbf{p}_3)^2 + (\mathbf{p}_2-\mathbf{p}_4)^2\right].
\ee

The order parameter $\Delta(\mathbf{r})$ in this expression should be written in terms of the eigenstates 
$\psi_n(\mathbf{r})$ of quadratic  part of the action $S_2$:
$\Delta(\mathbf{r})=\sum_n a_n \psi_n(\mathbf{r}).$
 As a result, a quartic mode--coupling between $a_{n}$ arises. 
We will first discuss the contributions where all  eigenstate indices are equal: 
$F_4^{(1)} =\frac{\nu}{2}\sum_n |a_n|^4 B_n$. It is optimal to rewrite $B_n$ as follows:
\bea
\label{Bn}
&B_n= \pi T\sum_{\epsilon}\int d^2r \left[|\epsilon| |G_{n,\epsilon}|^4(\mathbf{r})+\right.\\
&\left. \frac{D}{2} \left( |G_{n,\epsilon}|^2(\mathbf{r}) \Re G_{n,\epsilon}(\mathbf{r}) F_{n,\epsilon}^*(\mathbf{r})
 + \Re {\mathbf H}_{n,\epsilon}^2(\mathbf{r}) (G_{n,\epsilon}^*)^2(\mathbf{r})\right)\right]\nonumber
\eea
with  
\bea
\label{GHF}
G_{n,\epsilon}(r)=\sum_p \frac{e^{ipr}}{|\epsilon|+\frac12 D p^2}w_p(\epsilon)\Psi_n(\mathbf{p}),\\
{\mathbf H}_{n,\epsilon}(r)=\sum_p \frac{e^{ipr}i{\mathbf p}}{|\epsilon|+\frac12 D p^2}w_p(\epsilon)
\Psi_n(\mathbf{p}),\nonumber\\
F_{n,\epsilon}(r)=\sum_p \frac{e^{ipr}p^2}{|\epsilon|+\frac12 D p^2}w_p(\epsilon)\Psi_n(\mathbf{p}).\nonumber
\eea
where factors $w_q(\epsilon)$ are defined in Eq.~(\ref{wq2}) and $\Psi_n(\mathbf{p})$ stay for Fourier-transform
of eigenstates $\psi_n(\mathbf{r})$.
Below we evaluate integrals in Eqs.~(\ref{GHF}) at $T=0$
for eigenfunctions $\psi_n(\mathbf{r})$ localized at relatively short $L_n$.
These integrals are dominated by $\epsilon \sim D/L_n^2$ and  $p \sim 1/L_n$, 
thus the factor $w_q(\epsilon) \approx 1/\cosh(2\lambda_g\ln(L_n/l))$.
For  the final result of integration in Eq.~(\ref{Bn}) we find an estimate
\be
B_n \sim \frac{L_n^2}{D^2 \cosh^4(2\lambda_g\ln\frac{L_n}{l}) },
\label{Bn-f}
\ee
which differs  from the dimensional estimate $B_n \sim L_n^2/D^2$\, (provided originally in Ref.~\cite{SpivakZuzinHruska})
by $\cosh^4(...)$ factor only. We do not expect this modification to be significant due to smallness of $\lambda_g$
and not very large ratio $L_n/l$ for relevant localized eigenstates $\psi_n(\mathbf{r})$.

\subsection{Relevant time-scales of localized superconducting modes}

At large positive $\alpha_i = E_i + \delta_0$  typical frequency  of eigenmode is 
\begin{equation}
\label{wi}
\omega_i \sim \alpha_i/\lambda_g\Gamma_i \sim \alpha_i D/\lambda_g L_i^2.
\end{equation} This estimate comes from the comparison between
1st and 3rd terms in the action (\ref{action1}).  At large negative $\alpha_i $ the energy is minimized by
$|a_i|^2 = -\alpha_i/\lambda_g B_i \sim |\alpha_i| D^2/\lambda_g L_i^2 $  and two different fluctuation modes appear. Longitudinal mode
corresponds to variation of  $|a_i|$, and its frequency is $\omega_i \sim |\alpha_i| D/\lambda_g L_i^2 =
D |E_i + \delta_0|/\lambda_g L_i^2$. 
We will see now that at large enough $|\alpha_i|$,  the typical timescale of the transverse (phase rotation) mode
become  much longer.  We define local phase $\varphi_i(t) $ via relation  $a_i = |a_i| e^{i\varphi_i(t)}$ and
obtain phase-dependent action in the frequency domain
\begin{eqnarray}
\label{S-phi-omega1}
S[\varphi] = \frac{\nu}{\lambda_g} \int d\omega |\omega| \Gamma_i |a_i|^2 \left(\exp(i\varphi)\right)_\omega
 \left(\exp(-i\varphi)\right)_{-\omega} \\ 
= \tilde{c}\frac{g}{4\pi\lambda_g} \int d\omega |\omega| (\bar{\lambda} - E_i) \left(\exp(i\varphi)\right)_\omega
 \left(\exp(-i\varphi)\right)_{-\omega}
\label{S-phi-omega2}
\end{eqnarray}
In  Eq.~(\ref{S-phi-omega2}) we substituted the estimates for $\Gamma_i$ and $B_i$ together with the relation 
$4\pi\nu D = g$ and  $\tilde{c}$ is some factor $\sim 1$.
 Equivalent  action in the time domain is
\begin{equation}
S[\varphi(t)] = \frac{G_i}{2\pi^2} \int dt_1 dt_2\frac{\sin^2[(\varphi(t_1)-\varphi(t_2))/2]}{(t_1-t_2)^2},
\label{Sphi}
\end{equation}
where
\be
 G_i = \frac{g\tilde{c}}{\lambda_g} |E_i + \delta_0| =  - \frac{g\tilde{c}}{\lambda_g} ( E_i + \delta_0).
\label{G0}
\ee
The action (\ref{Sphi}) is similar to the one defined in
 Eq.~(3) of  Ref. \cite{FL1998}, where phase dynamics of artificially prepared superconducting islands was
studied; the constant  $G_i$ plays the role of the effective Andreev conductance measured in units of $4e^2/2\pi\hbar$.

Below we consider islands with large values of $G_i$,  which definitely exist due to large parameter $g \gg 1$.
For such islands,  autocorrelation function $C_i(t) = \langle\cos(\varphi(0)-\varphi(t))\rangle$ decreases
 logarithmically~\cite{FL1998} at moderate times $t \leq t_i$, while at the longest time scales 
$C(t) \propto G_i^{-1}(t_i/t)^2$,  where  correlation time of the $i$-th island 
\begin{equation}
t_i \approx \omega_i^{-1} \exp(G_i/2).
\label{ti}
\end{equation}
Below we focus on exponential dependence of $t_i$ on the parameters of islands entering (\ref{ti}) via $G_i$,
and neglect variations of prefactors $\omega_i = |\alpha_i| D/\lambda_g L_i^2$, 
replacing them by some typical frequency scale 
\be
\omega_\textrm{typ} =  \frac{\alpha_\textrm{typ} D}{\lambda_g L_\textrm{loc}^2}.
\label{omega1}
\ee
Here $\alpha_\textrm{typ}$ is the typical value of  $|E_i + \delta_0|$ for relevant islands, and $L_\textrm{loc}$ is their
typical localization length.
Exponential relation (\ref{ti}) together with exponential form of the DoS of localized states,
Eq.~(\ref{logdos}), lead to  the power-law tail in the probability distribution for the phase relaxation
rates $\gamma_i = 1/t_i$. Normalizing this probability distribution per unit area, and making use of Eqs.~(\ref{logdos},\ref{G0},\ref{ti},\ref{omega1}) we find, 
in the range $\gamma_i < \omega_\textrm{typ}$:
\begin{equation}
P_0(\gamma) d\gamma \approx 
\frac{p_0}{L_{\mathrm{loc}}^2} \left(\frac{\gamma}{\omega_\textrm{typ}}\right)^{\eta_0} \frac{d\gamma}{\omega_\textrm{typ}} =
\frac{p_0 \lambda_g }{D \alpha_\textrm{typ}} \left(\frac{\gamma}{\omega_\textrm{typ}}\right)^{\eta_0} d\gamma,
\label{P0gamma}
\end{equation}
using $L^2_{\mathrm{loc}}$ for typical area of relevant islands. Here
\bea
\label{eta-0}
 \eta_0 & = & \frac{2C_1(w)\lambda_g}{\tilde{c} g}-1,  \\ 
p_0(\delta_0)  & = &  2 \frac{\lambda_g}{\tilde{c} g} e^{-C_0(w)-C_1(w)\delta_0}
\label{p-0}
\eea
with $p_0 \ll 1 $ for the probability to find a superconducting island with phase relaxation rate $\gamma \sim \omega_\textrm{typ}$
within an area $\sim L_\textrm{loc}^2$.   This probability is exponentially low in the normal metal state,
 where $C_1(w)\delta_0 \gg 1$. Let us discuss how it is affected by variations in the main quantities describing the system. i) Growth of  average attraction $\bar{\lambda}$ leads to decrease of $\delta_0$ (see Eq.~(\ref{defd0})) and thus to sharp increase  of the density of islands with slow relaxation rates. ii) Increase of the disorder parameter $w$
diminishes the power-law exponent $\eta_0$, see Fig.\ref{fig:pg}c.  iii) Increase of the film conductance $g$ at \textit{fixed value} of $w$ would result in decrease if $\eta_0$. However, since $ w \propto 1/l^2 \propto 1/g^2$ (see discussion in the end of Sec.~\ref{sec:AL}) and $C_1(w)$ grows fast with decrease of $w$, the increase of $g$ translates to increase of $\eta_0$.

Exponent $\eta_0$ plays  crucial role in the further analysis.  The (extended) critical domain near
q--SMT  is characterized by $ \eta_0 \leq 1$,  while at  $\eta_0 \gg 1$  superconducting islands
are of little importance for the macroscopic properties of the film. Note that average correlation time
$\langle 1/\gamma \rangle$ is finite for $\eta_0 >0$, while its variance diverges 
for all $\eta_0 \leq 1$. We will now consider the interaction
between different islands and show that this interaction leads to renormalization of $\eta_0$ downwards.

\subsection{Inter-island coupling}

Now we proceed with the calculation of the interaction between order parameters of  distant localized islands $m$ and $n$, both unstable with respect to appearance of nonzero amplitudes of the order parameter, 
$a_{n} = |a_{n}|e^{i\varphi_{n}}$ and  $a_{m} = |a_{m}|e^{i\varphi_m}$.
Within the quadratic approximation  defined by the action $S_2(\Delta)$, Eq.~(\ref{S2}), 
these islands do not interact by construction:  $\psi_{n} (\mathbf{r})$ are the eigenfunctions of the
corresponding linear operator, Eqs.~(\ref{D1},\ref{E1}).   To derive the Josephson--type coupling $F_\textrm{int}(\varphi_n-\varphi_m)$
we need to account for non-Gaussian contributions to the action, so our starting point is given by
Eqs.~(\ref{F4},\ref{B4}).

\begin{figure}[tbp]
\includegraphics[width=0.5\textwidth]{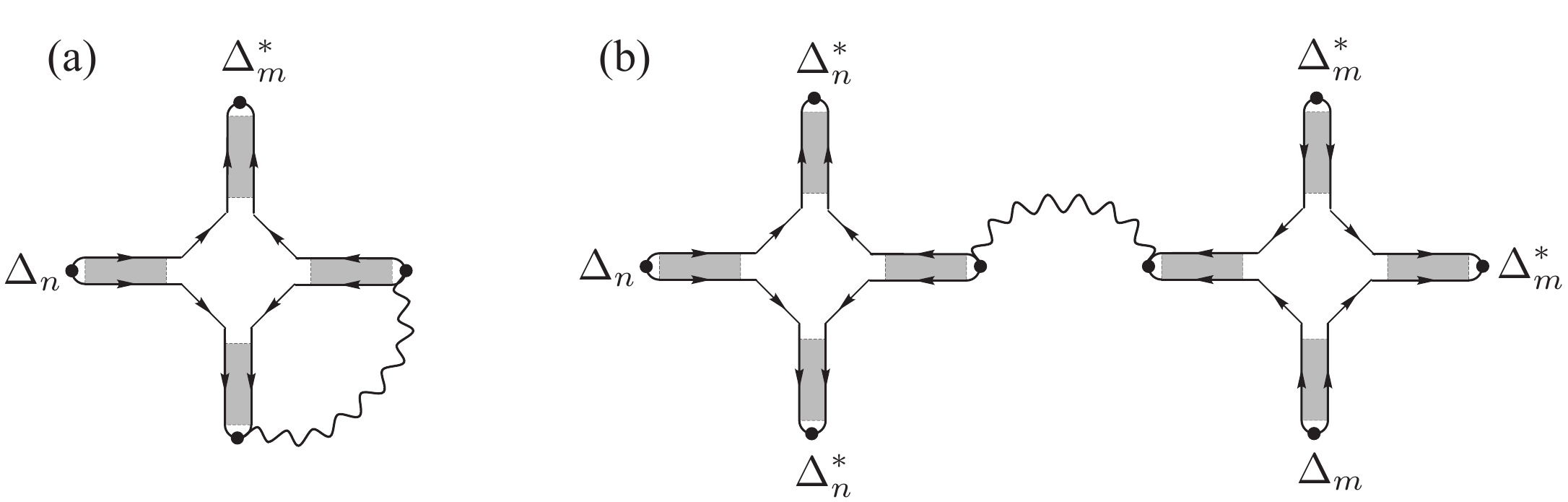}
\caption{First-order (a) and second-order (b) diagrams for the coupling between phases $\varphi_{n,m}$
 of distant islands. Wavy lines stay for the propagator $L_\omega(q)$.}
\label{Diagram2}
\end{figure}

The simplest relevant diagram is shown in Fig.~\ref{Diagram2}a, it contains one nonlinear vertex $B_4$
and one loop with dynamic superconducting propagator $L_\omega(q) $, see Eq.~(\ref{L2}).
We choose here $\Delta_n = a_n \psi_n(\mathbf{r})$ and $\Delta_m = a_m \psi_m(\mathbf{r})$ describing $n$--th and $m$--th localized islands, while summation over two extended modes is expressed via their propagator $L_\omega(q)$.
 For time-independent phases $\varphi_{n,m}$, 
the result of integration over frequency and momenta in the loop shown in Fig.~\ref{Diagram2}a can be written in the form of the interaction energy:
\be
 E_{\textrm{int},1}^{n,m} = - \tilde{J}_{nm} \cos(\varphi_m-\varphi_n),
\label{interE1}
\ee
where matrix elements $\tilde{J}_{nm} = \tilde{J}(R_{nm})$ depend on the distance 
$R_{nm}$ between centers of localized eigenstates $\psi_{n,m}(\mathbf{r})$; this distance is well defined
as long as $R_{nm} \gg L_\textrm{loc}$. An important result of the calculation  is that zero Fourier-harmonic of this
interaction, $\tilde{J}(Q=0) = \sum_{\mathbf{R}} \tilde{J}(\mathbf{R})$, does not contain any singularity when
the parameter $\delta_{\textrm{SCBA}}$  entering Eq.~(\ref{L2}) approaches zero.  In other terms, the integral defining
$\tilde{J}(Q=0)$  is determined by the ultraviolet region (large $q,\omega$) insensitive to the proximity
to a bulk superconductor-metal transition.  

Below we consider another source of long-range interaction,
 which does contain singular enhancement at $\delta_{\textrm{SCBA}} \to 0$ and thus
is the key driving force which establishes long-range coherence between well-separated islands. 

The most long-range contribution to the pair-wise interaction energy
 between phases  $\varphi_m $ and  $ \varphi_n $
appears in the second order  of perturbation theory over nonlinear vertex $B_4$  and contain one superconducting
propagator $L_\omega(\mathbf{R})$, with its Fourier representation in Eq.~(\ref{L2}), see Fig.~\ref{Diagram2}b.
In terms of the interaction contribution to the action $S_{\textrm{int}}$ we find, using Eqs.~(\ref{S4},\ref{F4},\ref{B4}):

\begin{widetext}
\begin{eqnarray}
\label{Sint}
S_{\textrm{int}}  =  - \frac{\nu^2}{4}\int dt dt' \int\Pi_{i=1}^4 d^2r_i \Pi_{i=1}^4 d^2r_i'
 B_4(\left\{\mathbf{r}_i\right\}) 
\Delta_n(\mathbf{r}_1,t)\Delta_n^*(\mathbf{r}_2,t)\Delta_n(\mathbf{r}_3,t) \\  \nonumber
\times B_4(\left\{\mathbf{r'}_i\right\}) \Delta_m^*(\mathbf{r}_1',t')\Delta_m(\mathbf{r}_2',t')\Delta_m^*(\mathbf{r}_3',t')
\times L(\mathbf{r}_4 - \mathbf{r}_4',t - t'),
\end{eqnarray}
\end{widetext}
where propagator $L (\mathbf{r}_4 - \mathbf{r}_4',t - t') =
\langle \Delta(\mathbf{r}_4,t)\Delta^*(\mathbf{r}'_4,t')\rangle $ describes quantum fluctuations of delocalized modes. 
 $\Delta_{n}(\mathbf{r}) = a_n\psi_n(\mathbf{r}) $ and 
$\Delta_{m}(\mathbf{r}')= a_m \psi_m(\mathbf{r}') $  correspond to the order parameters of 
superconducting islands $n,m$.
We assume the islands to be localized around their centers located at $\mathbf{R}_n$ and $\mathbf{R}_m$, and define
$\psi_n(\mathbf{r}) = \tilde{\psi}_n(\mathbf{r}- \mathbf{R}_n)$ and 
$\psi_m(\mathbf{r}') = \tilde{\psi}_m(\mathbf{r}'- \mathbf{R}_m)$.
 Here $\tilde{\psi} (\tilde{\mathbf{r}})$ functions are localized around zeroes of their arguments within 
small lengths $L_{n,m} \ll |\mathbf{R}_n-\mathbf{R}_m|$.  
It is convenient now to use space-time Fourier representation in terms of $L_\omega(q)$
for the propagator $L(\mathbf{r}_4 - \mathbf{r}_4',t - t')$, in order to split the whole multiple 
integral in Eq.~(\ref{Sint}) into two factorized parts:
\bea
\label{Sint2}
- S_{\textrm{int}}[\varphi_n(t),\varphi_m(t)]  = 
 \frac{\nu^2}{4} \,\, |a_n|^3 |a_m|^3 \,\,\,\,\,
   \\ \nonumber
\times \int \frac{d\omega}{2\pi} \frac{d^2q}{(2\pi)^2}  \Re \left[ s_n(\omega) s_m^*(\omega) A_n A_m^* \right]\, L_\omega(q) \, e^{i\mathbf{q}(\mathbf{R}_n - \mathbf{R}_m)},
\eea
where $s_{n,m}(\omega)$ stay for the Fourier-transforms of the time-domain functions $e^{i\varphi_{n,m}(t)}$.
Coefficients $A_n$ ($A_m$)  contain 4 integrals over space coordinates $\mathbf{r}_k$ ($\mathbf{r}_k'$).
Consider the structure of  $A_n$ more closely ($A_m$ is completely analogous):
\be
A_n = \Pi_{i=1}^4\int d^2r_i B_4(\left\{\mathbf{r}_i\right\}) 
\tilde\psi_n(\mathbf{r}_1) \tilde\psi_n^*(\mathbf{r}_2) \tilde\psi_n(\mathbf{r}_3)e^{i\mathbf{q}\mathbf{r}_4}.
\label{An0}
\ee
 The major dependence on $q$ in the final expression (\ref{Sint2}) 
comes from $L_\omega(q)$, and factors $A_n,A_m$ can be considered as $q$-independent.
Then the representation (\ref{B4}) for  $B_4(\left\{\mathbf{r}_i\right\})$
 can be used to rewrite $A_n$ in the form similar to Eq.~(\ref{Bn}) for $B_n$. At zero temperature:
\begin{widetext}
\bea
\label{A1}
A_n = \int_0^{1/\tau}\frac{(\epsilon\tau)^{\lambda_g} d\epsilon}{\epsilon}
\int d^2r \left[ \epsilon \tilde{G}^3_{n,\epsilon}(\mathbf{r}) + 
\frac{D}{4} \tilde{G}_{n,\epsilon}^*(\mathbf{r}) \mathbf{\tilde{H}}^2_{n,\epsilon}(\mathbf{r})\,\,\, 
+ \frac{D}{8} 
\left( 2 |\tilde{G}_{n,\epsilon}|^2(\mathbf{r}) \tilde{F}_{n,\epsilon}(\mathbf{r})
 + \tilde{G}^2_{n,\epsilon}(\mathbf{r}) \tilde{F}_{n,\epsilon}^*(\mathbf{r})
 \right)  
\right],
\eea
\end{widetext}
where  functions $\tilde{G},\tilde{F},\mathbf{\tilde{H}}$ are defined like functions
 $G,F,\mathbf{H}$ in Eqs.~(\ref{GHF}), but with the replacements
$\Psi_n(\mathbf{p}) \to \tilde\Psi_n(\mathbf{p})$.  Screening factors
 $w_p(\epsilon)=1/\cosh\left(2\lambda_g \ln(pl)\right)$, entering integrals in (\ref{GHF}), 
will be set to unity, since relevant  $ p \sim 1/L_n$ are relatively large and $\lambda_g \ll 1$.
The largest contribution to $A_n$
 comes from the last two terms in (\ref{A1}), which contain $\tilde{G}^2 \tilde{F}$ products.
Functions $\tilde{G}(\mathbf{r})$ and $\tilde{F}(\mathbf{r})$ are localized within the range
about $L_n$ around their maxima, while their amplitudes at maximum can be estimated as $L_n/D$ and
$1/(L_n D)$ correspondingly. The energy  integral in Eq.~(\ref{A1}) produces extra factor $1/\lambda_g$.
As a result, we come to the estimate  $A_n \sim \frac{L_n^3}{\lambda_g D^2}$.

 Taking also into account  $|a_n| \approx \sqrt{|\alpha_n|/\lambda_g}D/L_n$, we find eventually
\be
S_{\textrm{int}}[\varphi_n,\varphi_m]  = 
- \int \frac{d\omega}{2\pi} s_n(\omega) s_m^*(\omega) J_{nm}^{(\omega)}(\mathbf{R}_{nm})
\label{Sint3}
\ee
where (see Eq.~(\ref{L2}))
\be
J_{nm}^{(\omega)}(\mathbf{r}) \approx  \frac{g}{\lambda_g^4}  |\alpha_n|^{3/2} |\alpha_m|^{3/2}
\int \frac{D (d^2q/(2\pi)^2) e^{i\mathbf{q}\mathbf{r}}}{\delta_{\textrm{SCBA}} + 2 [(ql)^2 + 2|\omega|\tau]^{2\lambda_g}}
\label{J-omega}
\ee
Note that localization lengths $L_{n,m}$ cancel out from Eq.~(\ref{J-omega}), once we put
$\cosh(2\lambda_g\ln(L_n/l)) \approx 1$. Eqs.~(\ref{Sint3},\ref{J-omega}) define interaction 
action for two superconducting islands.  

Generally, this action cannot be reduced to
the interaction Hamiltonian, due to frequency dispersion entering $J_{nm}^{(\omega)}(\mathbf{r})$.
However, if one is interested in \textit{relatively slow} fluctuations of phases $\varphi_{n,m}(t)$, then
$s_{n,m}(\omega) \approx e^{i\phi_{n,m}}\delta(\omega)$ and one may use, instead of the action (\ref{Sint3}), 
the Hamiltonian of the form of Eq.~(\ref{interE1}), but with matrix elements
\be
J_{nm}(\mathbf{R}_{nm}) = A_{nm} \int \frac{d^2q}{(2\pi)^2} 
\frac{e^{i\mathbf{q}\mathbf{R}_{nm}}}{\delta_{\textrm{SCBA}} + 2 (ql)^{4\lambda_g}},
\label{Jnm}
\ee
with
\be
A_{nm} \propto \frac{\nu D^2}{\lambda_g^4}  |\alpha_n|^{3/2} |\alpha_m|^{3/2},
\label{Anm}
\ee
where we omitted numerical factor of the order of unity.
Below we will assume that reduction of the full dynamic
problem (\ref{Sint3})  to the Hamiltonian with matrix elements (\ref{Jnm}) is a good approximation.
We checked this assumption in the Supplement S1,  where  analysis of the two-island dynamics with full
interaction $ J_{nm}^{(\omega)}(\mathbf{r})$ was performed.

For the interaction strength in real space we obtain, after integration in Eq.~(\ref{Jnm}):
\begin{equation}
 J_{nm}(\mathbf{r}) \approx
  \frac{A_{nm}}{\pi}\frac{\lambda_g (r/l)^{4\lambda_g}}{r^2 [1 + (\delta_{\textrm{SCBA}}/2) (r/l)^{4\lambda_g}]^2}.
\label{J1}
\end{equation}
Note the presence of long (at small $\delta_{\textrm{SCBA}}$)  spatial scale $r^* = l \delta_{\textrm{SCBA}}^{-1/4\lambda_g} $.
Function $J_{nm}(r) \sim 1/r^{2-4\lambda_g} $  does not depend on $\delta_{\textrm{SCBA}}$  for $ r \leq r^*$, and it scales
as  $J_{nm}(r) \sim  \delta_{\textrm{SCBA}}^{-2} \times 1/r^{2+4\lambda_g} $  at  $ r \geq r^*$.  Major contribution to the
integral  $ \mathcal{J}_{nm}(0) = \int J(r) d^2 r \sim A_{nm}/\delta_{\textrm{SCBA}}$ comes from  $ r \sim r^*$.

Below we assume that typical interacting pairs of islands are separated by a large distance $r_{nm} \geq r^*$.
In result, the total interaction Hamiltonian reads
\be
H_{\textrm{int}} = - \sum_{nm}  J_{nm} \cos(\varphi_n-\varphi_m),
\label{Fint2}
\ee
where  coupling matrix elements
\be
J_{nm} = \frac{C_{nm}}{|(\mathbf{R}_n - \mathbf{R}_m)/l|^\beta}
\label{Jnm0}
\ee
and the coefficients $C_{nm}$ are: 
\be
C_{nm} = \frac{4\lambda_g A_{nm}}{\pi \delta_{\textrm{SCBA}}^2  l^2} \propto   
\frac{\nu v_F^2}{\lambda_g^3\delta_{\textrm{SCBA}}^2} |\alpha_n|^{3/2} |\alpha_m|^{3/2},
\label{Cnm}
\ee
where we omitted numerical factor of order unity. To obtain last form of $C_{nm}$, we used relation $D = v_F l/2$; for isotropic model of 2D
metal, $\nu v_F^2 = \epsilon_F/\pi$.
The constants $C_{nm}$ contain  product of large factors $\epsilon_F$ and 
$\delta_{\textrm{SCBA}}^{-2}$ by small $\sim \alpha^3_{n,m}$.  It is  assumed that typical distance 
$R_{nm} = |\mathbf{R}_n - \mathbf{R}_m| $ is much longer than $l$.

Our analysis below is based upon the single-island action (\ref{Sphi}) 
and the interaction Hamiltonian (\ref{Fint2}).


\section{Effect of  interaction between  islands at $T=0$:  strong disorder RG }
\label{sec:RG}

\subsection{General approach and RG equations}

Results of Sec. \ref{sec:AL}  demonstrate crucial property of the q--SMT  in presence of sufficiently
strong fluctuations $\delta\lambda(\mathbf{r})$:  locally superconducting regions (islands)
appears at random locations and are localized  on typical length $L_{\mathrm{loc}}$  which is much shorter
than typical distance between islands 
\be
L_0  \sim \frac{L_{\mathrm{loc}}}{\sqrt{p_0} } \gg L_{\mathrm{loc}},
\label{L00}
\ee
where $p_0$ is defined in Eq.~(\ref{p-0}).
Far from the SMT, in the normal state, the density $n_{\mathrm{isl}} = 1/L_0^2$ 
is too low for interaction terms, Eq.~(\ref{Fint2})  to be relevant, thus phases of individual
islands fluctuate independently.   With increase of average attraction and decrease of $\delta_0$,
the density $n_{\mathrm{isl}}$ grows, as well as the  probability that some  islands occur to be close 
enough to interact strongly. For each pair of islands, $n$ and $m$,
the strength of this interaction $J_{nm}$, defined in Eq.~(\ref{Jnm0}),  is to be compared to 
the relaxation rates of the same islands $\gamma_{n},\gamma_{m}$. 
We are interested especially in the range of parameters  where the key exponent $\eta_0$ satisfies $0< \eta_0 \leq 1$,
thus individual decay rates $\gamma_n$ are distributed widely  in the small--$\gamma$ domain.

Quantum fluctuations of  phases $\varphi_n(t)$ and $\varphi_m(t)$ are mutually independent if  at least one of the rates
$\gamma_n$, $\gamma_m$  is much larger than the coupling energy $J_{nm}$.  If, however, 
$\min(\gamma_n,\gamma_m) \ll J_{nm}$, then dynamics of phases $\varphi_n(t), \varphi_m(t)$
 becomes correlated at the time scales longer than $1/J_{nm}$ and phase
difference $\varphi_n(t) - \varphi_m(t)$ ceases to grow with time.  
 In result (see Sec. S1 of the Supplemental Material for details), the joint two-island decay
rate $\gamma_{nm}$ becomes much smaller than the individual rates $\gamma_n$ and $\gamma_m$:
\be
\gamma_{nm} = \frac{\gamma_n \gamma_m}{J_{nm}} \quad \mathrm{for} 
\quad \min(\ln(\gamma_n),\ln(\gamma_m)) < \ln(J_{nm}).
\label{fusion1}
\ee
To understand Eq.~(\ref{fusion1}) it is enough to notice that lowest-frequency Andreev conductance of the
two-island system $G_{nm} = G_n + G_m$, while the relevant pre-exponential factor in the expression like
Eq.~(\ref{ti}) is given now by the timescale $1/J_{nm}$ where interaction between islands sets in; 
see Eq.~(\ref{W1long}) for details.

`Fusion rule'  (\ref{fusion1}) shows that interaction between different islands slows down dynamics of their phases.
Eventually, under many such fusions, it can lead to the complete freezing of this dynamics, 
leading to macroscopic phase coherence and superconductivity.
An appropriate quantitative method to describe this phenomenon is known as 
Strong Disorder Renormalization Group (SDRG) developed originally for one-dimensional quantum Ising model in random
transverse field~\cite{Fisher92,Fisher95}  and employed later on for numerous different problems, see 
reviews~\cite{Igloi-Monthus05,Refael13,Igloi-Monthus14}. The SDRG method is useful when 
the Hamiltonian contains competing terms with random amplitudes, and  one  (or more) of these amplitudes
is characterized by a very broad probability distribution. Precisely this property  invalidates the use of
mean-field approach~\cite{FL1998,SLF2001}, even in the case of long-range interactions, 
like proximity coupling (\ref{Jnm0}). The idea of SDRG is to integrate out  quantum degrees of 
freedom of large system sequentially, starting from the highest energy scale, and to derive 
 stochastic evolution equations for the remaining amplitudes entering  the Hamiltonian.

The most similar physical problem treated by this kind of approach~\cite{Hoyos07,Hoyos09,Maestro08,Maestro10}, 
refers to the q--SMT in quasi-one-dimensional wires with strong pair-breaking.  Dynamics of individual islands
was considered to be of the same kind as we described above, see Eqs.~(\ref{ti},\ref{fusion1}).
However, in these references the proximity coupling was considered to be short-ranged (due to strong pair-breaking), and the problem was thus reduced to a nearest-neighbor coupling model. It results~\cite{Hoyos07}  in  a multiplicative recursion relations for renormalized couplings $J_{ij}$, similar 
in its structure to the one for individual decay rates, Eq.~(\ref{fusion1}).
Then the whole SDRG  belongs to the same universality class as random-field Ising model~\cite{Fisher92,Fisher95}.
This is not the case for our problem with long-range proximity coupling (\ref{Jnm0}),
with $2 < \beta  < 3$.  However,  SDRG scheme appropriate for our model was also developed and studied, 
although in somewhat different contexts ~\cite{Igloi14,Altman04}. 
Below we adapt to our problem the method of Ref.~\cite{Igloi14} which is a dual alternative of the one developed 
in Ref.~\cite{Altman04}.

We employ a 2D version of the `primary model' defined in Ref.~\cite{Igloi14}, where a problem of a quantum 
transverse-field Ising model with power-law exchange coupling of the type (\ref{Jnm0}) was studied.
Interaction between islands starts to become relevant when the magnitude of $J_{nm}$,  Eqs.~(\ref{Jnm0},\ref{Cnm}), 
for typical  nearest-neighbor distance $R_{nm} \sim L_0$ becomes comparable to the typical value of island's
relaxation rates $\omega_\textrm{typ}$ defined in Eq.~(\ref{omega1}). The corresponding condition can be written, with the use
of Eqs.~(\ref{omega1},\ref{Jnm0},\ref{Cnm}),
in the form
\be
[p_0(\delta_0)]^{\beta/2} \propto \left(\frac{\delta_{\textrm{SCBA}}}{\alpha_\textrm{typ} g }\right)^2,
\label{p0crit}
\ee
where we omitted numerical factor of order unity. Here  $\alpha_\textrm{typ}$ is the  magnitude of $|E_i + \delta_0|$ for typical relevant islands. In the further analysis
we put $\alpha_\textrm{typ} \approx 1/g$, in order to get moderately large $G_i$ for relevant islands, see Eq.~(\ref{G0}).
While deriving Eq.~(\ref{p0crit}) we put  $(L_\textrm{loc}/l)^{4\lambda_g} \approx 1$, as it was already done earlier.

In order to find the value of $\delta_0$ entering Eq.~(\ref{p0crit}), we need to solve  Eqs.~(\ref{p-0},\ref{p0crit}) 
together. As a result, we find:
\be
\delta_{\textrm{SCBA}} \propto \frac{1}{g^{3\beta/8}} \exp\left[-\frac{\beta}{4}(C_0+C_1\delta_c)\right]  \ll 1,
\label{delta1}
\ee
where we omitted numerical factor of order unity. 
The corresponding length-scale
$L_0$ is given by
\be
L_0 \sim  L_{\mathrm{loc}}\, g^{3/4} \, \exp\left[\frac{1}{2}(C_0+C_1\delta_c)\right] \gg L_{\mathrm{loc}}.
\label{L01}
\ee 
The relation (\ref{L01}) determines the spatial length scale $L_0$, where the SDRG procedure starts in. The corresponding energy scale $\Omega_0 = \omega_\textrm{typ}$ is defined in Eq.~(\ref{omega1}). 

We now describe briefly the SDRG approach.  Starting from the upper energy cutoff $\Omega_0 \sim \omega_\textrm{typ}$\,
(see Eq.~(\ref{omega1})),  
at any value of running RG energy scale $\Omega \leq \Omega_0$,  we look for the largest energy parameter
 in the system, it is equal to $\Omega$ by definition. It can be either the rate $\gamma_i$ of the $i$-th island,
 or the coupling $J_{nm}$ between $n,m$ pair of them. In the first case, $i$-th island is decimated,
while couplings between remaining islands are left unchanged.  As  a result, typical area corresponding 
to distance between nearest remaining islands grows linearly:  $S_{nm} = S_{ni} + S_{im}$.  The same
relation can be written in the form 
$J_{nm}^{-2/\beta} = J_{im}^{-2/\beta} + J_{ni}^{-2/\beta}$.
Now we introduce dimensionless variable $y = (\Omega/J)^{2/\beta} - 1$, so
 it vanishes at $J=\Omega$. 
 The above recursion relation for $J_{nm}$ reads then as
\be
y_{nm}   = y_{ni} + y_{im} + 1.
\label{y-rec}
\ee
Below we will see that actual probability distribution $Q(y)$ becomes very broad near the q--SMT, thus 
the term $1$ in the R.H.S. of Eq.~(\ref{y-rec}) can be neglected.

If the largest  energy parameter in the Hamiltonian close in its value  to
 $\Omega$ is some coupling $J_{nm}$, its decimation leads to modification
of the rates $\gamma_{n,m}$ according to Eq.~(\ref{fusion1}).  It is convinient to introduce logarithmic
variables $x_n = (2/\beta) \ln(\Omega/\gamma_n)$, then the   recursion relation equivalent to (\ref{fusion1})
reads as
\be
x_{nm} = x_n + x_m.
\label{x-rec}
\ee
Factor $2/\beta$ in the definition of $x_n$ variable is introduced in order to simplify the following
equations, since the same factor enters the definition of $y_n$.

Derivation of the functional RG equations for probability densities $P(x)$ and $Q(y)$ corresponding to stochastic
equations Eqs.~(\ref{y-rec},\ref{x-rec}) is provided  in Refs.~\cite{Altman04,Refael13}. We reproduce it in Supplement S2
together with some extension.
The RG evolution parameter is defined by
the logarithmic variable $\tau = (2/\beta) \ln(\Omega_0/\Omega)$  which counts logarithm
of average area per a cluster in the system of interacting islands with a largest energy scale $\Omega$.
The functional RG equations allow quasi-stationary solutions of  exponential form:
\be
P(x,\tau)=p(\tau)e^{-p(\tau)\,x}\, \qquad Q(y,\tau)=q(\tau)e^{-q(\tau)\, y}
\label{distrib}
\ee  
if the functions $p(\tau)$ and $q(\tau)$ obey the system of equations
\bea
\label{RG1}
\frac{dp}{d\tau}= - p\cdot q, \\ 
\frac{d q}{d\tau} = - p\cdot q + q.
\label{RG2}
\eea
Equations (\ref{RG1},\ref{RG2}) are formally equivalent to the Kosterlitz RG equations~\cite{Kosterlitz74} 
for 2D XY model.  They differ from similar equations of the Fisher's SDRG by the presence of the
 last term $q$ in (\ref{RG2}).
This term is due to trivial scaling dimension $1$  of the variable $y$, while variable $x$ is dimensionless.

The system of equations (\ref{RG1},\ref{RG2}) possesses the first integral $q(\tau) - p(\tau) + \ln p(\tau) = \mathrm{Const}$ which allows to reduce it to a single equation for $p(\tau)$. Below we will be most interested in the
vicinity of the critical point where $p(\tau)$ is close to unity. Thus we denote $p(\tau) = 1 + \xi(\tau)$,
 keep major terms of expansion over $\xi(\tau) \ll 1$ (we will find that $\xi\sim\sqrt{\delta}$) and obtain single RG equation
\be
\frac{d\xi}{d\tau} = - \frac{\xi^2}{2} + \delta,
\label{RGsingle}
\ee
where $\delta \ll 1$ parametrizes the distance to the critical point. Within the same accuracy,
$q(\tau) = \xi^2/2 - \delta$.  

During the RG transformation, SC islands continuously merge, so the areal density $n(\tau)$ of survived islands 
decreases according to the equation
\begin{equation}
\frac{d n}{d\tau} = - \left(p(\tau) + q(\tau)\right) n(\tau),
\label{n-t}
\end{equation}
which follows from the fact that at each decimation  (island or bond) the number of surviving 
islands decreases by one. Its solution is
\begin{equation}
n(\tau_\Omega) = n_0 \exp\left(-\int_0^{\tau_\Omega} \left(p(\tau)+q(\tau)\right)\right) d\tau,
\label{n-r}
\end{equation}
where $n_0 \sim L_0^{-2}$ is the initial density of islands at the starting energy scale of RG, $\Omega_0$.

Another important characteristic of the inhomogeneous state formed due to RG procedure is  average number 
of islands $N(\tau_\Omega)$ which constitute a cluster formed at the scale 
\be
\label{deftauO0}
\tau_\Omega = (2/\beta)\ln(\Omega_0/\Omega).
\ee
Contrary to $n(\tau_\Omega)$, this quantity is determined by the integration over the whole RG trajectory.
We present necessary calculations in the Section S2 of the Supplemental Material. The result is given by
\begin{equation}
 N(\tau_\Omega) = \int_0^{\tau_\Omega} p_1(\tau) d\tau,
\label{N-r}
\end{equation}
here function $p_1(\tau) \equiv p_1(\xi(\tau))$, where $p_1(\xi)$  solves Eq.~(\ref{p1xi}). 
Explicit solutions following from Eqs.~(\ref{n-r}) and (\ref{N-r}) will be presented below, separately for $\delta >0$ and $\delta < 0$.

\subsection{Metal phase: line of RG fixed points}

At small  $\delta > 0$,  full solution
to Eqs.~(\ref{RG1},\ref{RG2}) reads:
\bea
\label{sol1}
 p(\tau) = 1 + \sqrt{2\delta} \coth\left(\sqrt{\frac{\delta}{2}}(\tau + \tau_+)\right), \\
 q(\tau) = \frac{\delta}{\sinh^2\left(\sqrt{\frac{\delta}{2}}(\tau + \tau_+)\right)}.
\label{sol2}
\eea
Integration constant $\tau_+$ and key parameter $\delta$ should be determined by the 
matching of quasi-stationary distributions
(\ref{distrib}) to the bare distributions  $P_0(\gamma)$ (see Eq.~(\ref{P0gamma})) and $Q_0(J)$:
\bea
\label{tau+1}
1+ \sqrt{2\delta} \coth\left(\tau_+\sqrt{\frac{\delta}{2}}\right) = \frac{\beta}{2} (1 + \eta_0), \\
\frac{\delta}{\sinh^2\left(\tau_+ \sqrt{\frac{\delta}{2}}\right)} = Q_0(J \sim \Omega_0).
\label{tau+2}
\eea
To derive Eq.~(\ref{tau+1}), we employed the relation $\gamma \propto \exp(-\beta x/2)$. Right-hand sides of
both Eqs.~(\ref{tau+1},\ref{tau+2}) are of the order of unity; thus we need  $\tau_+ \sim 1$ 
in order that $\delta \ll 1 $ cancels out from the LHS's of these equations.  Then we see that
at $\delta \ll 1$  the integration constant $\tau_+$ is not important for the asymptotic solutions
and will be ignored below. 

The solution  provided  by Eqs.~(\ref{sol1},\ref{sol2}) defines a line of fixed points parametrized
by $\delta >0 $.  In the infrared limit $\tau \to \infty$,  
we find 
$p(\infty) = 1+ \sqrt{\delta/2}$, and  $q(\tau) \approx  4\delta e^{-\sqrt{2\delta}\tau }$.
The scale $\tau_\delta=1/\sqrt{2\delta}$ marks the end of 
renormalization process. The corresponding energy and spatial scales read:
\be
\Omega_\delta = \Omega_0 \, e^{-\frac{\beta}{2\sqrt{2\delta}}}\,, \qquad 
L_\delta = L_0 \, e^{\frac{1}{2\sqrt{2\delta}}}.
\label{corr1}
\ee
At $L \gg L_\delta$  interaction between islands is too weak to change
the distribution of the relaxation rates, while the ratio of the decimation rates $q(\tau)/p(\tau)$
starts to drop fast with $L$.  Density of surviving islands  drops with decrease of $\Omega$, according to
Eqs.~(\ref{n-r},\ref{sol2}) as (see Eq.~(\ref{deftauO0})):
\begin{equation}
n_+(\tau_\Omega) = n_0 \left(\frac{\Omega}{\Omega_0}\right)^{2/\beta}
\frac{\delta}{\sinh^2(\frac{\sqrt{2\delta}}{\beta}\ln\frac{\Omega_0}{\Omega})}.
\label{n-r-1}
\end{equation}
For the average number of islands inside a cluster, $N(\tau)$, we use Eqs.~(\ref{N+S})  to obtain
\begin{eqnarray}
\label{N2}
N_+(\tau_\Omega) = \frac{3}{2\delta} \left[ \tau_\Omega \sqrt{\frac{\delta}{2}} 
\coth\left(\tau_\Omega\sqrt{\frac{\delta}{2}}\right)  - 1  \right] \approx
\\ \nonumber
\left\{
\begin{array}{rl}
\frac14\tau^2_\Omega \qquad  \tau_\Omega \ll \delta^{-1/2} \\
\frac{3}{2\sqrt{2\delta}} \tau_\Omega \qquad \tau_\Omega  \gg \delta^{-1/2}
\end{array}
\right.
\end{eqnarray}
First line above corresponds to the close vicinity of the critical line, and the result
 $N_+(\tau) \propto \tau^2$ coincides with the one present in Ref.~\cite{Igloi14}.
The second line demonstrates that logarithmic growth of $N_+(\tau_\Omega)$ continues even at arbitrary low energy scales,
where $\xi(\tau_\Omega)$ saturates and both $q(\tau_\Omega)$ and $n_+(\tau_\Omega)$ vanish.

We do not determine here the dependence of the key RG parameter  $\delta$ on the original parameters 
like $\delta_0$ and $g$.  This dependence may occur to be nontrivial, and we leave this question for future studies.

\subsection{Superconducting phase: slow runaway of the RG flow}

At negative values of $\delta$, the RG equation (\ref{RGsingle}) has qualitatively different
solution
\bea
\label{sol3}
\xi(\tau) = \sqrt{2|\delta|} \cot\left(\sqrt{\frac{|\delta|}{2}}(\tau+\tau_-)\right),  \\
q(\tau) = \frac{|\delta|}{\sin^2\left(\sqrt{\frac{|\delta|}{2}}(\tau+\tau_-)\right)}.
\label{sol4}
\eea
Integration constant $\tau_- \sim 1 $ can be ignored at $|\delta| \ll 1 $  for the same reason as 
described above for $\tau_+$. The solution (\ref{sol3}) for $\xi(\tau)$ changes sign at  $\tau= \tau_0 \approx \pi/\sqrt{2|\delta|}$, while
$q(\tau)$, has a minimum at the same $\tau$.  In a broad vicinity of  $\tau_0$, the function
$q(\tau)$ is nearly constant, which translates to a weak temperature dependence of $q_T$ in a broad range
of low temperatures.

Density of surviving islands  behaves now as
\begin{equation}
n_{-}(\tau_\Omega) = n_0 \left(\frac{\Omega}{\Omega_0}\right)^{2/\beta}
\frac{\delta}{\sin^2(\frac{\sqrt{2\delta}}{\beta}\ln\frac{\Omega_0}{\Omega})}.
\label{n-r-2}
\end{equation}
For the average number of islands $N_-(\tau_\Omega)$ in a typical cluster, we use Eq.~(\ref{N-S}):
\begin{equation}
 N_{-}(\tau_\Omega) = \frac{3}{2|\delta|}  \left[
1 - \tau_\Omega\sqrt{\frac{|\delta|}{2}}  \cot\left( \tau_\Omega\sqrt{\frac{|\delta|}{2}}  \right)
 \right].
\label{N3}
\end{equation}
At large $\tau > \tau_0$ superconducting correlations
(measured by $q(\tau)$) start to grow, and $p(\tau) = 1 + \xi(\tau)$ decreases.  Near the point $\tau = 2\tau_0$
the solution (\ref{sol3},\ref{sol4}) develops a singularity. 
Average number of islands in a typical cluster, $N_{-}(\tau)$, also diverges as $\tau \to 2\tau_0$. 

Physically, this solution corresponds to emergence of a globally coherent superconducting state with  typical value of the zero-temperature
order parameter 
\be
 \Delta_\delta(0) = \Omega_0 \, e^{-\frac{\beta \pi}{\sqrt{2|\delta|}}} \equiv
\Omega_0 \,\left(\frac{\Omega_\delta}{\Omega_0}\right)^{2\pi}.
\label{Delta}
\ee
Note  that $\Delta_\delta$  is parametrically smaller than $\Omega_\delta$ defined by Eq.~(\ref{corr1}) 
at the same value of $|\delta|$, due to the presence of extra $2\pi$ in the exponent in Eq.~(\ref{Delta}).
Such a state is very fragile,  with extremely low transition temperature, $T_c(\delta) \sim \Delta_\delta$.

Upon approach to  the  point $\tau_\Omega = 2\tau_0$, the RG equations (\ref{RG1},\ref{RG2}) and their consequences
(\ref{n-r-2}) and (\ref{N3}) loose their applicability.
It happens when the inequality $\xi(\tau) \ll 1$ is not valid anymore. For very small $|\delta| \ll 1$ it corresponds to
 $\tau$ rather close to $2\tau_0$, see Eq.~(\ref{sol3}).


\section{Strange metal: qualitative discussion.}
\label{sec:SM}

\begin{figure}[tbp]
\includegraphics[width=0.4\textwidth]{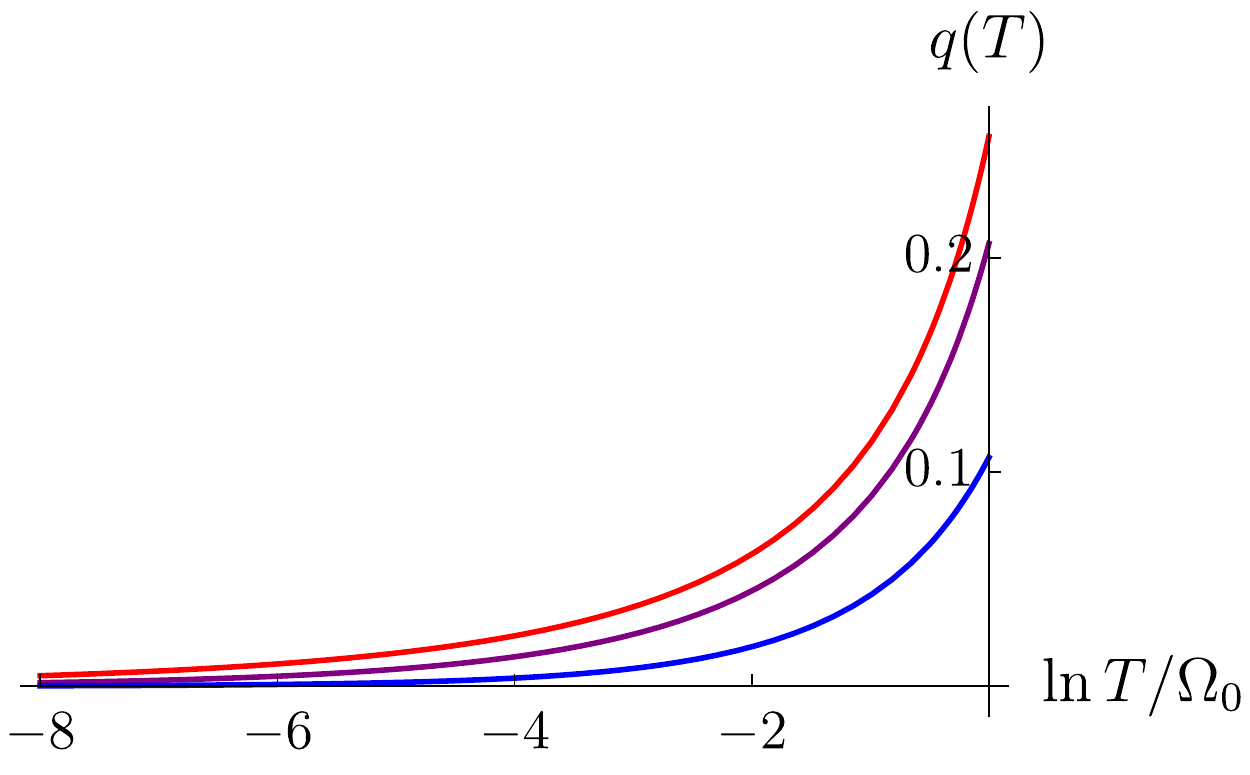}
\caption{Function $q(T)$ as found from Eqs.~(\ref{RG1}), (\ref{RG2})} in the normal phase: $\delta =0.05, 0.1, 0.2$ from red to blue.
\label{fign}
\end{figure}

The range of initial parameters leading  to the solution, Eqs.~(\ref{sol1}),~(\ref{sol2}), of the
strong-disorder RG equation, corresponds to the ground-state of the Griffiths type, 
which  is not globally superconducting. It contains clusters of superconducting islands those phases are locked in together
by proximity couplings. Sizes of these clusters  vary in a broad range, up to the correlation length 
$L_\delta$  given by Eq.~(\ref{corr1}). Typical number of original superconducting islands in the largest clusters
is $N_\delta \sim 1/\delta $, see Eq.~(\ref{N2}).

At any $T>0$, strong-disorder RG flow of Sec.~\ref{sec:RG}  should be stopped at the lowest energy scale
$\Omega_T= T$, corresponding to $\tau_T = (2/\beta)\ln(\Omega_0/T)$.  Interactions of smaller  magnitudes,
$J \leq T$, do not lead to any noticeable phase correlations between clusters which were 
formed at higher energy scales, i.e. at $\tau < \tau_T$.  At the final stage of RG, distribution functions (\ref{distrib}) which 
are formed at $\tau=\tau_T$, can be rewritten in terms of energy variables $J$ and $\gamma$
as follows:
\bea
\label{PJ}
\mathcal{P}(\gamma;T) d \gamma & = & 
\frac{2\,p_T}{\beta}\left(\frac{\gamma}{T}\right)^{\frac{2 p_T}{\beta} - 1} \frac{d\gamma}{T}
\\
\mathcal{Q}(J;T) dJ & = & 
\frac{2\,q_T}{\beta} \left(\frac{T}{J}\right)^{\frac2{\beta}} 
e^{- q_T\left(\frac{T}{{J}}\right)^{2/\beta}}
 \frac{dJ}{J},
\label{Qg}
\eea
where $(\gamma,J) \leq T$. Close to the critical point $p_T \approx 1 + \sqrt{2\delta}$ and
$q_T \ll 1$ is determined by Eq.~(\ref{sol2}) with $\tau = \tau_T$.  
Eqs.~(\ref{PJ},\ref{Qg}) demonstrate that typical value of  phase relaxation rate $\gamma(T) \sim T$.
Typical  inter-cluster coupling energy  $J(T) \sim T q_T^{\beta/2} \ll T$, 
demonstrating  weakness of interaction between largest clusters.   

Observe that Eq.~(\ref{PJ}) shows that for any $\beta > 2$, the exponent 
$ \frac{2 p_T}{\beta} $  always becomes smaller than unity for small enough $\delta$, i.e. close enough
to  the q--SMT critical point (but before the transition point is reached). As soon as $ \frac{2 p_T}{\beta} < 1$, the average correlation time
 $\langle 1/\gamma \rangle =  \int_0^T\mathcal{P}(\gamma;T)d\gamma/\gamma $
diverges, while $ \langle \gamma \rangle \sim T$. This divergence does not imply global phase coherence, but rather indicates strong fluctuations of relaxation rates between different clusters. Indeed, correlation time $t^{(a)}$ of any $a$--th cluster is bounded from above by its classical value 
\be
 t_a^{cl} =  \frac{G_a \hbar}{2\pi T},
\label{cl-t}
\ee
which can be found from the action (\ref{Sphi}) estimating phase diffusion at $T >0$:
\be
 \frac12 \langle(\phi_a(0)-\phi_a(t))^2\rangle \approx 
\frac{2\pi}{G_a} \frac{T |t|}{\hbar}.
\label{thermal}
\ee
Here $G_a$  stands for the total Andreev conductance (in units of $4e^2/h$)
between the $a-$th cluster of strongly coupled superconducting islands and surrounding normal 
metal.  Since we found previously that Andreev conductances of individual islands $G_i$ sum up during
their `merging' under RG procedure, we expect $G_a$ to be proportional to the number of islands $N_a$
 constituting $a$--th cluster.  Therefore $G_a^{\mathrm max} \sim 1/\delta$,  as follows from Eq.~(\ref{N2})
at $\Omega \sim \Omega_\delta$. In result, the longest phase correlation time of largest coupled 
clusters is estimated as
\be
t_{\mathrm{max}}(T) \sim \frac{\hbar}{2\pi T}\frac1{\delta}.
\label{t-max}
\ee
Note that $ T t_{\mathrm{max}}(T)$ is singular as $\delta$ goes to zero, while typical relaxation times 
$t_\textrm{typ} \sim \hbar/T$.
This is one of the specific feature of the Griffiths phase characterized by the presence of  arbitrary large fractal clusters.

The presence of very large but mutually uncorrelated clusters, leads to an important contribution to the electron
dephasing rate $\tau_\varphi^{-1}$.  Contribution of individual islands to dephasing rate was analyzed in Ref.~\cite{Skvortsov04}.
Crucial feature of its main result, Eq.~(3),  is as follows: Andreev reflection  contribution to 
$\tau_\varphi^{-1}$ is nearly $T$-independent at $T$ much above macroscopic $T_c$. The same feature is expected to hold
for the Griffiths state with large mutually incoherent clusters.

\begin{figure}[tbp]
\includegraphics[width=0.4\textwidth]{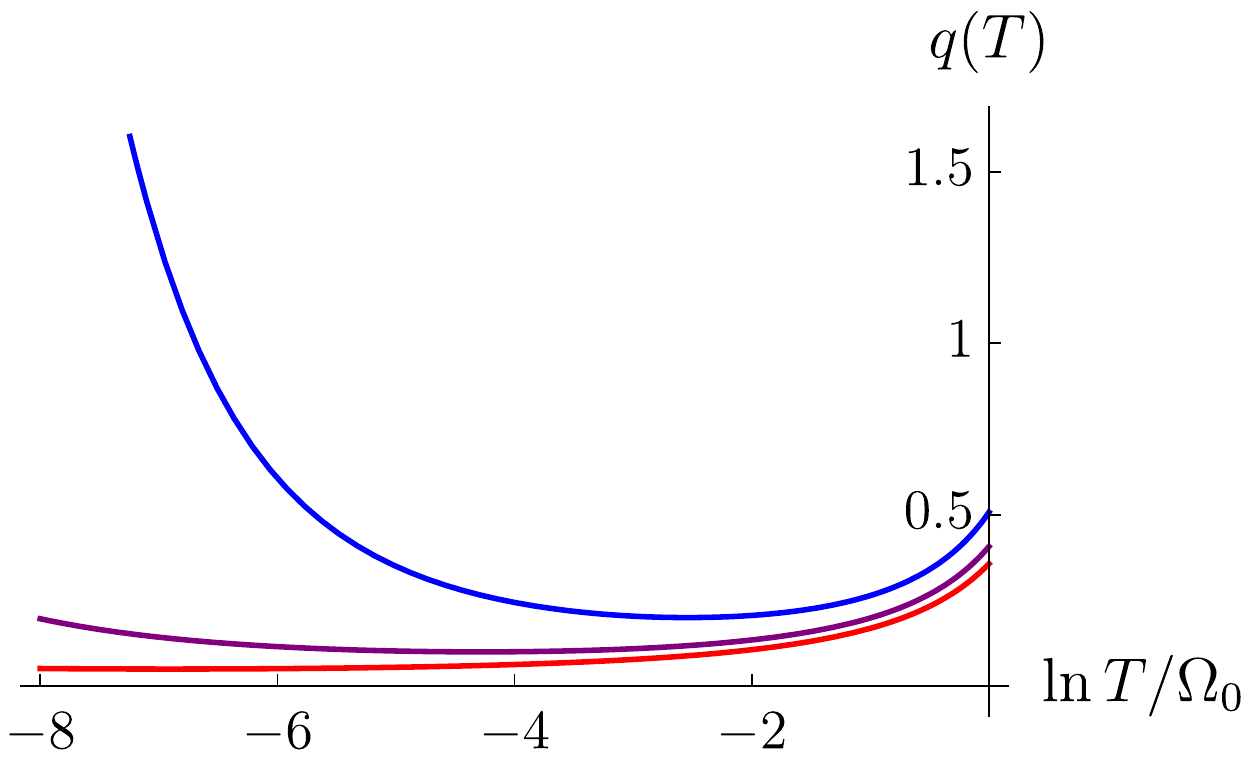}
\caption{Function $q(T)$ as found from Eqs.~(\ref{RG1}), (\ref{RG2})} in the superconducting phase: $-\delta =0.05, 0.1, 0.2$ from red to blue.
\label{figsc}
\end{figure}

`Superconducting' side of the quantum phase transition is described by Eqs.~(\ref{sol3},\ref{sol4}) and demonstrates a very
unusual feature:  due to the presence of extremum of $q(\tau)$  at $\tau_T = \tau_0(\delta) = \pi/\sqrt{2|\delta|}$,  
temperature dependencies of physical quantities are expected to be very weak in a broad temperature range
\be
 T_c = \Omega_0 e^{- \frac{\pi\beta}{\sqrt{2|\delta|}} }   \ll  T  \ll  \Omega_0 e^{-\frac{\beta}{2\sqrt{2|\delta|}}}.
\label{interval}
\ee
The relative extension of this range grows enormously as $|\delta|$ decreases, see Fig.~\ref{figsc}.  
Typical number of islands within largest clusters
$N_a \geq 1/|\delta|$ is the same  or larger  (see Eq. (\ref{N3})) than in the Griffiths metal phase, 
and the above conclusion about  nearly--$T$--independent contribution to the dephasing rate 
is applicable in the whole interval (\ref{interval}).

We believe that  region of  phase diagram with small negative $\delta$, while being formally superconducting, 
is a good candidate for the description of a `strange metal' state, for the reasons described in the above paragraph.
Note that approximate $T$--independence (in the sense described above)  comes together with a strong sensitivity to magnetic
field:  typical scale of magnetic field, which corresponds to a single flux quantum per relevant area, 
$L_{\delta_0}^2 = L_0^2 e^{\tau_0}$, is
\be
H_\delta = \frac{\Phi_0}{L_0^2} \exp\left( - \frac{\pi}{\sqrt{2|\delta|}}\right).
\label{H0}
\ee
At $B \geq H_\delta$ renormalization flow is modified as $\tau$ reaches $\tau(B) = \ln(\Phi_0/B L_0^2) < \tau_0(\delta)$, 
since proximity coupling becomes frustrated at longer length-scales.
 In the close vicinity of the critical point $\delta=0$ relevant range
of magnetic field becomes unexpectedly small.   The magnetic field--controlled quantum transition  
has been studied before \cite{Hoyos07,Hoyos09,Maestro08} in the assumption that proximity coupling is short--ranged due to random magnetic frustration. However, at $T=0$  exponential decay of proximity coupling refers to the disorder-averaged $\langle E_J(r) \rangle$ only, while its second
 moment is still described~\cite{Spivak1995} by a power-law function of distance $r$.

We note that large spatial dimensions of weakly-coupled superconducting clusters  
makes the system near the critical point unusually susceptible to a weak \textit{rf}-frequency noise,
like the one demonstrated in Ref.~\cite{Tamir19}. 


\section{Conclusions}
\label{sec:Conclu}

We demonstrated  breakdown  of   scaling theory~\cite{BK} of  quantum Superconductor-Metal transition in 
thin films, due to spontaneous formation of  localized islands of superconductivity.  The latter is shown to be
 a generic consequence of  sufficiently strong spatial fluctuations of the Cooper attraction strength 
$\delta\lambda(\mathbf{r})$ (these fluctuations always become strong~\cite{Skvortsov2005}
close to the mean-field q--SMT via Finkel'stein mechanism~\cite{Finkelstein1987,Finkelstein1994}). This unusual scenario, with a disorder threshold for an appearance of localized Lifshits tail in the density of locally superconducting regions, 
is realized when effective Cooper-channel repulsion constant $\lambda_g = 1/\sqrt{2\pi g} < 1/4 $, corresponding
to a normal-state resistivity of a film $R_{\square} \leq 10$ KOhm. We have also found that the effective strength of fluctuations $\delta\lambda(\mathbf{r})$ increases with decrease of the electronic mean-free-path $l$. 

Power-law interaction, see Eq.~(\ref{Jnm0}),
 between phases of different emergent islands, together with power-law distribution
of their individual relaxation rates, Eq.~(\ref{P0gamma}), 
lead to formation a Griffiths-type phase in the vicinity of a genuine
transition to superconducting state.  This transition is described by a version of Strong-Disorder Renormalization
Group, formally similar to the one employed in Refs.~\cite{Altman04,Igloi14}. Metallic Griffiths phase
is described as a line of fixed points of this RG.  Physically this phase consists of large fractal clusters
of superconducting islands, strongly coupled to each other.  The largest size of such clusters is given
by $L_\delta$ in Eq.~(\ref{corr1}), while the number of individual islands in large clusters scales as $1/\delta$,
where  $\delta$ is  the \textit{renormalized}  distance to the quantum critical point.
Note an important feature of these large fractal clusters: Andreev conductance between such a cluster and surrounding
metal is proportional to the number of islands $N_\delta$  it consists of,  due to parallel nature of Andreev reflection
processes which occur at different islands.

The relation between renormalized parameter $\delta$ and \textit{bare} distance to the critical point 
$\delta_\textrm{SCBA}$ is not yet established; it can be nonlinear. In order to find this relation, one needs to find exact relations
between microscopic parameters  $\delta_0$ and $g$, and initial conditions $p(0), q(0)$ for the SDRG equations
(\ref{RG1},\ref{RG2}). We leave this interesting problem for future studies.

The most unusual observation of this paper is related to the superconducting phase realized at $\delta < 0$.
Namely, non-monotonic character of the RG flow in this region of the phase diagram leads to a very weak 
temperature-dependence of physical properties (including  dephasing rate) in a broad range of temperatures, 
see Eq.~(\ref{interval}) and Fig.~\ref{figsc}.  It might provide a clue to understanding a `strange metal' phase near q--SMT transition.
Below $T_c$, a superconducting state occurs. Due to its strong spatial 
inhomogeneity, we expect it to be gapless, for the reasons understood originally in Ref.~\cite{LO72}.

We are grateful to A. V. Andreev, A. S. Ioselevich, V. E. Kravtsov, M. A. Skvortsov, B. Z. Spivak and 
S. V. Syzranov for useful discussions. 
This research was partially supported by the Skoltech NGP grant and by the Russian Academy of Sciences program
`Modern problems of low-temperature physics'. The research of KT was supported by the Russian Scientific Foundation, Grant No. 17-72-30036.


\bibliography{lambda}

\vspace{1cm}

\onecolumngrid

\begin{center}
{ \bf \Large Supplemental Material}
\end{center}

\setcounter{figure}{0}
\makeatletter
\renewcommand{\thefigure}{S\@arabic\c@figure}
\makeatother

\vskip1cm

{\bf\large Section S1:  Time-resolved slow dynamics of the two-island system}
\label{sec:S1}
We consider here the system of two small superconducting islands in contact with 2D metal, with imaginary-time 
action ($T=0$):
\be
S[\varphi_i(t)] = \int\int dt_1 dt_2 
\left[\sum_{i=1,2} \frac{G_i}{2\pi^2} \frac{\sin^2[(\varphi(t_1)-\varphi(t_2))/2]}{(t_1-t_2)^2}
    - \mathcal{J}_{12}(t_1-t_2) \cos\left(\varphi_1(t_1) - \varphi_2(t_2)\right)
\right],
\label{S3}
\ee
where $G_{1,2}$ are Andreev conductances of both islands, and  $\mathcal{J}_{12}(t)$ is  (compare with Eq.~(\ref{Jnm})):
\be
\mathcal{J}_{12}(t) = \int \int \frac{d\omega}{2\pi}\frac{d^2q}{(2\pi)^2}
\frac{A_{12}\,\, e^{-i\omega t + i \mathbf{q r}} }{ \delta_{\textrm{SCBA}} + 2 ( 2|\omega\tau| + (ql)^2 )^{2\lambda_g} }
\label{J12}
\ee
and $r = |\mathbf{r}_1-\mathbf{r}_2|$ is the distance between the islands.
The integral (\ref{J12}) defines a time-dispersive proximity coupling.  In the main text we neglected this
dispersion assuming our  system of interacting islands can be described in terms of a Hamiltonian, 
with matrix elements $J_{nm} = \int d t \mathcal{J}_{nm}(t)$.  Here  we take into account this time-dispersion
in the explicit form and derive the fusion rule  (\ref{fusion1}), which is one of our basic points for the SDRG analysis.
Below we will need the value of coupling strength in the frequency representation, $J(\omega)$, defined as follows:
\be
J(\omega) = \int\frac{d^2q}{(2\pi)^2}
\frac{A_{12}\,\, e^{ i \mathbf{q r}} }{ \delta_{\textrm{SCBA}} + 2 ( 2|\omega\tau| + (ql)^2 )^{2\lambda_g} }.
\label{Jomega}
\ee

We first consider Gaussian fluctuations of phases described by the action (\ref{S3}).  Slow component of the phase
difference $\varphi_1-\varphi_2$ will be denoted as  $\Phi$.  Expanding over fast components $\delta\phi_{1,2}$
up to second order, we find
\be
S_{\phi} = \frac12 \int\frac{d\omega}{2\pi} \delta\phi^T(-\omega) \hat{C}(\omega) \delta\phi(\omega),
\label{Sdeltaphi}
\ee
where
\be
\hat{C}(\omega) =     
\begin{pmatrix}
\frac{G_1}{2\pi}|\omega| + J(0)\cos\Phi & - J(\omega)\cos\Phi \\
- J(\omega)\cos\Phi & \frac{G_2}{2\pi}|\omega| + J(0)\cos\Phi 
\end{pmatrix}
 \equiv
\begin{pmatrix}
X_1(\omega) & - Y(\omega) \\
- Y(\omega) & X_2(\omega) 
\end{pmatrix}.
\label{Comega}
\ee
We introduce three correlation functions: $ W_1(t)= \frac12 \langle (\delta\phi_{1}(0) - \delta\phi_1(t))^2 \rangle$,
$ W_2(t)= \frac12 \langle (\delta\phi_{2}(0) - \delta\phi_2(t))^2 \rangle$, and 
$W_{12}(t) = \frac12 \langle (\delta\phi_1(0) - \delta\phi_2(t))^2 \rangle$. For a vector 
$\left( W_{1}(t),W_2(t) \right)$ we find:
\be
\left( W_1(t),W_2(t) \right)  = 
\int\frac{d\omega}{2\pi}\frac{\left( X_2(\omega), X_1(\omega) \right)}
{X_1(\omega) X_2(\omega) - Y^2(\omega)}\left(1-e^{-i\omega t}\right),
\label{W1}
\ee
while
\be
W_{12}(t) = \int\frac{d\omega}{2\pi}
\frac{\frac12(X_1(\omega)+X_2(\omega)) - Y(\omega) e^{-i\omega t} }{X_1(\omega) X_2(\omega) - Y^2(\omega) }.
\label{W12}
\ee
First we consider two limiting cases for the integrals (\ref{W1},\ref{W12}) : 
short-time asymptotics $ t \ll 1/J(0)$ and long-time one for $t \gg 1/J(0)$.  In the first case
we can neglect all terms with $J$ coupling in Eq.~(\ref{W1}) and obtain
\be
W_{1,2}(t \ll 1/J(0)) = \frac{2}{G_{1,2}}\ln(\omega_0 t),
\label{W1short}
\ee
where $\omega_0$ is the high-frequency cut-off for the action (\ref{S3}).

In the opposite limit $t \gg 1/J(0)$, the $t$-dependent contribution to the  integral (\ref{W1})
comes from lowest-$\omega$ region
\be
W_1\left(t \gg \frac1{J(0)}\right) = \int\frac{d\omega}{2\pi}\frac{[(G_2/2\pi)|\omega| + J(0)\cos\Phi ] (1- e^{-i\omega t})}
{\frac{G_1G_2}{4\pi^2}\omega^2 + \frac{G_1+G_2}{2\pi}|\omega| J(0)\cos\Phi + (J^2(0)-J^2(\omega))\cos^2\Phi}
\approx \frac{2  \ln(J(0) t) }{G_1 + G_2} + \frac{2}{G_1}\ln\frac{\omega_0}{J(0)}.
\label{W1long}
\ee
Here we keep  in numerator the term $J(0)\cos\Phi$ which is the largest at $\omega \to 0$.
In denominator the first term is $\propto \omega^2$ and can be neglected in comparison with
second term, which is  linear in $\omega$. The last term is also $\propto \omega$ but it is much less than
the second one:
$$
\frac{J^2(0)-J^2(\omega)}{J(0)\, |\omega|} \sim  \frac{J(0)\, r^2}{D} \ll 1.
$$
As a result, the smallest-frequency region $\omega \leq J(0)$
 in the integral (\ref{W1long}) leads to the first term in the R.H.S. of (\ref{W1long}).  
The second term  comes from 
 larger-frequency range $J(0) \ll \omega \ll \omega_0$.  

Finally, we note that the difference $W_1(t) - W_{12}(t)$ does not diverge in the limit $ t \to \infty$:
\be
W_{12}(t)-W_1(t) = \int\frac{d\omega}{2\pi}
\frac{\frac1{4\pi}(G_1-G_2)|\omega| + [\frac{G_2}{2\pi}|\omega| + (J(0)-J(\omega)) \cos\Phi ] e^{-i\omega t}}
{\frac{G_1G_2}{4\pi^2}\omega^2 + \frac{G_1+G_2}{2\pi}|\omega| J(0)\cos\Phi + (J^2(0)-J^2(\omega))\cos^2\Phi }.
\label{W112}
\ee
Both numerator and denominator in the above integral are proportional to $\omega$ at $\omega \to 0$,
 thus the result of integration is finite.
It means that $\varphi_1$ and $\varphi_2$ are coupled and fluctuate together at the longest time-scales.
The strength of these fluctuations is given by the R.H.S. of Eq.~(\ref{W1long}).

\vspace{1cm}

{\bf\large Section S2:  Strong-disorder renormalization group equations}

Below we reproduce for our case derivation of the functional renormalization group equations of the strong-disorder 
type~\cite{Altman04,Refael13,Igloi-Monthus14}, with some appropriate extension which allows us to find evolution
of the typical number of islands $N(\tau)$ in a cluster.

Starting from stochastic recursion equations (\ref{y-rec},\ref{x-rec}) and following the logics of Ref.~\cite{Altman04}
one finds RG equations for distribution functions $P(x,\tau)$ and $Q(y,\tau)$ in the form:
SDRG equations read 
\begin{eqnarray}
\label{SDRG1}
\frac{\partial P(x)}{\partial\tau} = \frac{\partial P}{\partial x} +
Q(0) \int_0^\infty dx_1 P(x_1)P(x-x_1) + P(x) (P(0) - Q(0)), \\ 
\frac{\partial Q(x)}{\partial\tau} = (1+y)\frac{\partial Q}{\partial y} +
P(0) \int_0^\infty d y_1 Q(y_1) Q( y - y_1 - 1) + Q(y) (Q(0) - P(0) + 1).
\label{SDRG2}
\end{eqnarray}
Below we will find that relevant values of $y$ are very large (in other words, major part of the
distribution function corresponds to $y \gg 1$), which allows us to neglect the term $-1$ in the argument
of $Q(y-y_1-1)$ in Eq.~(\ref{SDRG2}).  After this simplification, Eqs.~(\ref{SDRG1},\ref{SDRG2}) allow
for the solution in the exponential form (\ref{distrib}), if the functions $p(\tau)$ and $q(\tau)$ solve
 Eqs.~(\ref{RG1},\ref{RG2}).

We will need also  more general distribution function $\mathcal{P}(x,N,\tau)$ which depends, 
in addition to $x$, on the number $N$ of islands which constitute a cluster. 
RG equation for this function can be obtained in the same way as it is done in~\cite{Igloi2002}.
Namely, consider the integral term in Eq.~(\ref{SDRG1})
which describes fusion of two clusters with rates $\gamma_1$ and $\gamma-\gamma_1$,
due to eliminating of the strong bond connecting them.
These clusters contain $N_1$ and $N-N_1$ islands, thus integral term in the equation for $\mathcal{P}(x,N,\tau)$
contains, in addition to the convolution over $y_1$, also the convolution over variable $N_1$:
\begin{equation}
\frac{\partial \mathcal{P}(x,N,\tau)}{\partial\tau} =
\frac{\partial \mathcal{P}}{\partial x} +
Q(0) \int_0^\infty dx_1 \mathcal{P}(x_1,N_1)\mathcal{P}(x-x_1,N-N_1) + \mathcal{P}(x,N) (P(0) - Q(0)).
\label{SDRGN}
\end{equation}
The structure of Eq.~(\ref{SDRGN}) allows us to simplify it
by  introducing the Laplace transform, $P_s(x,\tau) = \int_0^\infty \mathcal{P}(x,N,\tau)e^{-s N} dN$:
\begin{equation}
\frac{\partial P_s(x)}{\partial\tau} = \frac{\partial P_s}{\partial x} +
Q(0) \int_0^\infty dx_1 P_s(x_1)P_s(x-x_1) + P_s(x) (P(0) - Q(0)).
\label{Ps1}
\end{equation}
Solution of Eq.~(\ref{Ps1}) can be found in the form 
\begin{equation}
P_s(x,\tau) = \pi(s,\tau)e^{-r(s,\tau)}.
\label{Anz2}
\end{equation}
At $s=0$ the solution goes back to the known one for $P(x,\tau)$, therefore $\pi(0,\tau)=r(0,\tau)= p(\tau)$.

Substitution of the Anzats (\ref{Anz2})  to Eq.~(\ref{Ps1}) leads to the system of equations:
\begin{eqnarray}
\frac{\partial\pi}{\partial\tau} & = & ( p - q - r) \pi, \\ \nonumber
\frac{\partial r}{\partial\tau} & = &  - q \pi.
\label{pir}
\end{eqnarray} 
Using (\ref{Anz2}) it is easy to show that average number of islands in  clusters formed at the RG scale $\tau)$
is given by $N(\tau)=  p^{-1}(\tau)\frac{\partial (r- \pi)}{\partial s}|_{s=0}$.  Thus we expand functions
$\pi(s,\tau)$ and $r(s,\tau)$ up to the linear order
in $s$ : $\pi(s,\tau) = p(\tau) - s \pi_1(\tau)$ and  $r(s,\tau) = p(\tau) - s p_1(\tau)$.
New functions $p_1(\tau)$ and $\pi_1(\tau)$ obey the equations:
\begin{eqnarray}
\label{pir1}
\frac{d\pi_1}{d\tau} & = & - p p_1 - q \pi_1, \\ \nonumber
\frac{d p_1}{d\tau}  & = & - q \pi_1.
\end{eqnarray}
Remember that we solve our basic RG equations assuming $\xi(\tau) = p(\tau) - 1 \ll 1$.  We use the same
approximation here and exclude $\tau$ from the system of equations (\ref{pir1}) in favor of independent 
variable $\xi$. The result can be written in the form
\begin{equation}
\frac{d^2 p_1}{d\xi^2}  - \frac{2 p_1}{\xi^2 - 2\delta} = 0.
\label{p1xi}
\end{equation}
We neglected the term $- \frac{d p_1}{d\xi}$ in the above equation, since it is small at $\xi \ll 1$.
Subtracting 2nd of Eqs.~(\ref{pir1}) from the first one (and employing approximation $p(\tau) \approx 1$),
 we find that
\begin{equation}
 \frac{d{N}(\tau)}{d\tau} = \frac{d}{d\tau}\left(\pi_1(\tau) - p_1(\tau)\right) \approx p_1(\tau).
\label{Ntau}
\end{equation}
Therefore, to find $N(\tau)$ we need just to solve Eq.~(\ref{p1xi}), substitute $\xi $ for $\xi(\tau)$ in the solution
and integrate over $\tau$.

For $\delta >0$ we set $2\delta = a^2$. Physically acceptable solution of Eq.~(\ref{p1xi}) is given by
\begin{equation}
 p_1^{(+)}(\xi(\tau)) = \frac{3\xi}{2a^2} - \frac{3(a^2-\xi^2)}{4a^3}\ln\frac{\xi-a}{\xi+a} 
\equiv \frac3{2a}\left(\coth\frac{a\tau}{2} -\frac{a\tau/2}{\sinh^2\frac{a\tau}{2}}\right).
\label{p11}
\end{equation}
while $\xi(\tau) \equiv p(\tau) - 1$ is present in Eq.~(\ref{sol1}).
Solution (\ref{p11}) is normalized in such a way that $\lim_{a \to 0}p^{(+)}(\xi) = 1/\xi$.
With Eqs.~(\ref{p11}) and (\ref{Ntau}) we find
\be
N_+(\tau) = \frac{3}{a}\left( \frac{a\tau}{2}\coth\frac{a\tau}{2} - 1 \right).
\label{N+S}
\ee

For negative $\delta$ we put $-2\delta= a^2$, here $\xi(\tau) = a\cot\frac{a\tau}{2}$.
We find the solution for $p_1(\xi)$ in the form
\begin{equation}
p_1^{(-)}(\xi) = \frac{3}{2} \frac{\xi^2 + a^2}{a^2}\left(\frac{\pi}{2} - \arctan\frac{\xi}{a}\right)  - \frac{3\xi}{2a^2}
\equiv \frac3{2a}\left(\cot\frac{a\tau}{2} -\frac{a\tau/2}{\sin^2\frac{a\tau}{2}}\right).
\label{p12}
\end{equation}
Again, the solution (\ref{p12}) is normalized by condition $\lim_{a \to 0}p^{(-)}(\xi) = 1/\xi$. 
Using Eqs.~(\ref{Ntau}) we find
\be
N_-(\tau) = \frac{3}{a}\left(1 - \frac{a\tau}{2}\cot\frac{a\tau}{2} \right).
\label{N-S}
\ee

\end{document}